\documentclass[12pt]{article}
\usepackage{epsfig,amsmath,amssymb}
\usepackage{graphicx,psfrag} 
\numberwithin{equation}{section}

\newcommand{\geff}{g_{\mbox{\tiny eff}}}
\newcommand{\be}{\begin{equation}}
\newcommand{\ee}{\end{equation}}
\newcommand{\bea}{\begin{eqnarray}}
\newcommand{\eea}{\end{eqnarray}}
\newcommand{\bA}{\begin{array}}
\newcommand{\eA}{\end{array}}
\newcommand{\bc}{\begin{center}}
\newcommand{\ec}{\end{center}}

\newcommand{\ra}{\rightarrow}
\newcommand{\del}{\partial}

\newcommand{\ie}{{\it i.e.}}
\newcommand{\eg}{{\it e.g.}}

\newcommand{\scriptA}{\mathcal{A}}
\newcommand{\scriptF}{\mathcal{F}}

\begin{document}


\begin{titlepage}

\bc

\hfill 
\\         [22mm]

{\Huge Notes on hyperscaling violating Lifshitz\\ 
[2mm]
and shear diffusion}
\vspace{16mm}

Kedar S. Kolekar, Debangshu Mukherjee, K. Narayan\\
\vspace{3mm}
{\small \it Chennai Mathematical Institute, \\ }
{\small \it SIPCOT IT Park, Siruseri 603103, India.\\ }

\ec
\medskip
\vspace{30mm}

\begin{abstract}
We explore in greater detail our investigations of
shear diffusion in hyperscaling violating Lifshitz theories in
arXiv:1604.05092 [hep-th]. This adapts and generalizes the
membrane-paradigm-like analysis of Kovtun, Son and Starinets for shear
gravitational perturbations in the near horizon region given certain
self-consistent approximations, leading to the shear diffusion
constant on an appropriately defined stretched horizon. In theories
containing a gauge field, some of the metric perturbations mix with
some of the gauge field perturbations and the above analysis is
somewhat more complicated. We find a similar near-horizon analysis can
be obtained in terms of new field variables involving a linear
combination of the metric and the gauge field perturbation resulting
in a corresponding diffusion equation. Thereby as before, for theories
with Lifshitz and hyperscaling violating exponents $z, \theta$
satisfying $z<4-\theta$ in four bulk dimensions, our analysis here
results in a similar expression for the shear diffusion constant with
power-law scaling with temperature suggesting universal behaviour in
relation to the viscosity bound. For $z=4-\theta$, we find logarithmic 
behaviour.
\end{abstract}

\end{titlepage}

\newpage 
{\footnotesize
\begin{tableofcontents}
\end{tableofcontents}}

\section{Introduction}

In \cite{Kolekar:2016pnr}, we had studied the shear diffusion constant
in certain hyperscaling violating Lifshitz theories by obtaining it as
the coefficient of the diffusion equation satisfied by certain near
horizon metric perturbations. In the present paper, we explore this in
greater detail and study generalizations.

To put this in context, let us recall nonrelativistic holography or
gauge/gravity duality \cite{AdSCFT} which has been under active
exploration over the last few years. In particular, spacetimes
conformal to Lifshitz \cite{Kachru:2008yh,Taylor:2008tg}, referred to
as hyperscaling violating spacetimes arise in effective
Einstein-Maxwell-Dilaton theories \eg\
\cite{Gubser:2009qt,Cadoni:2009xm,Goldstein:2009cv,Charmousis:2010zz,
  Perlmutter:2010qu,Gouteraux:2011ce,Iizuka:2011hg,Ogawa:2011bz,
  Huijse:2011ef,Dong:2012se,Alishahiha:2012cm,Bhattacharya:2012zu}.
Null $x^+$-reductions of $AdS$ plane waves
\cite{Narayan:2012hk,Narayan:2012wn}, which are large boost, low
temperature limits \cite{Singh:2012un} of boosted black branes
\cite{Maldacena:2008wh} provide certain gauge/string realizations of
these. See \eg\
\cite{Dong:2012se,Balasubramanian:2010uk,Donos:2010tu,Ross:2011gu,
Christensen:2013rfa,Chemissany:2014xsa,Hartong:2015wxa,Taylor:2015glc} 
for aspects of Lifshitz and hyperscaling violating holography. Some of
these exhibit novel scaling for entanglement entropy \eg\
\cite{Ogawa:2011bz,Huijse:2011ef,Dong:2012se}, with the string
realizations above reflecting this
\cite{Narayan:2012ks,Narayan:2013qga,Mukherjee:2014gia,Narayan:2015lka},
suggesting corresponding regimes in the gauge theory duals exhibiting
this scaling.

Understanding hydrodynamic behaviour in these nonrelativistic
gauge/gravity dualities is of great interest: see \eg\
\cite{Pang:2009wa,Cremonini:2011ej,Kiritsis:2012ta,Hoyos:2013qna,Sadeghi:2014zia,
Roychowdhury:2014lta,Kiritsis:2015doa,Kuang:2015mlf,Blake:2016wvh,
Ling:2016ien,Davison:2016auk,Hartong:2016nyx,Patel:2016ymd} for 
previous and recent investigations.  Our approach in
\cite{Kolekar:2016pnr} to studying hydrodynamics and viscosity has
been somewhat different, and based on Kovtun, Son, Starinets
\cite{Kovtun:2003wp}. They observed that metric perturbations
governing diffusive shear and charge modes in the near horizon region
of the dual black branes of relevance simplify allowing a systematic
expansion. This results in a diffusion equation for these shear modes
on a stretched horizon, with universal behaviour for the diffusion
constant, thereby leading to the viscosity bound
\cite{Kovtun:2004de}. This is akin to the membrane paradigm
\cite{Thorne:1986iy} for black branes, the horizon exhibiting
diffusive properties. This approach is based simply on the fact that
near horizon metric perturbations lead to a diffusion equation: thus
it does not rely on any holographic duality per se. It is of course
consistent with holographic results \eg\
\cite{Policastro:2001yc,Son:2002sd}\ (see \eg\ \cite{Son:2009zzd} for
a review of these aspects of hydrodynamics).

In \cite{Kolekar:2016pnr}, we adapted the membrane-paradigm-like
analysis of \cite{Kovtun:2003wp} and studied the shear diffusion
constant in bulk $(d+1)$-dimensional hyperscaling violating theories
(\ref{hvmetric}) with $z,\theta$ exponents. Specifically the diffusion
of shear gravitational modes on a stretched horizon is mapped to
charge diffusion in an auxiliary theory obtained by compactifying one
of the $d_i$ boundary spatial dimensions exhibiting translation
invariance. This gives a near horizon expansion for perturbations with
modifications involving $z,\theta$. For generic exponents with
$d-z-\theta>-1$, we found the shear diffusion constant to be\ 
${\cal D} = {r_0^{z-2}\over d-z-\theta-1}$~, \ie\ power-law scaling
(\ref{DT(2-z)/z}) with the temperature $T\sim r_0^z$. Studying various
special cases motivated the guess (\ref{eta/s,DT(2-z)/zHV}), \ie\ 
$\# {\cal D} T^{{2-z\over z}} = {1\over 4\pi}$\ where $\#$ is some
$(d,z,\theta)$-dependent constant, suggesting that ${\eta\over s}$ has
universal behaviour. The condition $z<2+d_i-\theta$ representing this
universal sector appears related to requiring standard quantization
from the point of view of holography. 
When the exponents satisfy\ $d-z-\theta= -1$, the diffusion constant
exhibits logarithmic behaviour, suggesting a breakdown of
some sort in this analysis. The exponents arising in null reductions
of AdS plane waves or highly boosted black branes
\cite{Narayan:2012hk,Singh:2012un,Narayan:2012wn} mentioned above
satisfy this condition, which can be written as $z=2+d_{eff}$.

The analysis above arose solely from perturbations in the metric
sector. In theories with a gauge field, the near-horizon diffusion
equation analysis above must be extended to also include the gauge
field sector which mixes with some of the metric perturbations.  The
resulting story is somewhat more intricate, both calculationally and
conceptually, and is the subject of this paper. To give a flavour of
this, it is worth describing the analysis above in a little more
detail. Shear gravitational perturbations $h_{xy}, h_{ty}$, satisfy
the diffusion equation in the near-horizon region within certain
approximations, as stated earlier: they are mapped to $U(1)$ gauge
field modes $\scriptA_x, \scriptA_t$ upon compactifying the
$y$-direction which enjoys translation invariance. Near horizon
membrane currents can be appropriately defined in terms of the field
strengths for this gauge field $\scriptA_\mu=(\scriptA_t,\scriptA_x)$,
which then can be shown to satisfy Fick's law\ $j^x =
-\mathcal{D}\del_xj^t$, which in turn using current conservation leads
to the diffusion equation $\del_tj^t =\mathcal{D}\del_x^2j^t$, valid
within a self-consistent set of approximations imposed near
horizon. In terms of the original linearized Einstein equations for
metric perturbations (without this $y$-compactification), the
diffusion equation stems from one of the Einstein equations, which is
essentially a conservation equation schematically of the form\ $\del_x
(\del_r (\# h_{xy})) \sim \# \del_t (\del_r (\# h_{ty}))$ where the
$\#$ are $r$-dependent factors. The other linearized Einstein
equations are coupled second order equations for $h_{ty}, h_{xy}$. In
the case where the hyperscaling violating Lifshitz theory has a
background gauge field $A_\mu$, it turns out that the $h_{ty}$ metric
perturbation mixes with the gauge field component $a_y$. The resulting
Einstein equations along with the gauge field equation are coupled
equations for $h_{xy}, h_{ty}, a_y$\ (with the other modes decoupling
for modes respecting the $y$-compactification ansatz), and at first
sight they do not reveal any such diffusion-equation-type structure.

Towards understanding this better, it is important to note that the
hyperscaling violating Lifshitz black branes here are not charged
black branes: the gauge field and scalar here simply serve as sources
that support the nonrelativistic metric as a solution to the gravity
theory. Using intuition from the fluid-gravity correspondence
\cite{Bhattacharyya:2008jc}, the fact that these are uncharged black
branes means that the near-horizon perturbations are effectively
characterized simply by local temperature and velocity
fluctuations. Thus since charge cannot enter as an extra variable
characterizing the near-horizon region, the structure of the diffusion
equation and the diffusion constant should not be dramatically altered
by the presence of the gauge field.  In light of this intuition, a
closer look reveals that the relevant component of the Einstein
equation is of the form\ $\del_x (\del_r (\# h_{xy})) \sim \# \del_t
(\del_r (\# h_{ty})) - \del_t (\# a_y)$. This naively suggests that
perhaps the correct field variable in terms of which the Einstein
equation can be recast as a diffusion equation is in fact ${\tilde
  h_{ty}}\equiv h_{ty}-\int\# a_y dr$.\ Analyzing this in greater
detail shows that this essential logic is consistent, and thereby
leads to a generalization of the analysis in \cite{Kolekar:2016pnr}
mapping shear diffusion to charge diffusion after
$y$-compactification. This results in the same expression for the
shear diffusion constant but obtained using the leading near-horizon
expressions for ${\tilde h_{xy}} \equiv h_{xy}$ and ${\tilde h_{ty}}
\equiv h_{ty}-\int\# a_y dr$.

In sec.~2, we briefly review the results of \cite{Kolekar:2016pnr}
obtained by the $y$-compactification. Sec.~3 discusses this analysis
from the point of view of the original Einstein equations without
$y$-compactification, giving some insight into how the diffusion
equation effectively arises. In Sec.~4, we discuss the perturbations
in the general hyperscaling violating Lifshitz background
incorporating the gauge field perturbations as well. We then describe
the various modifications in terms of the new field variables leading
to the diffusion equation and thereby the shear diffusion constant.
Sec.~5 has a Discussion. The Appendices provide various technical
details.

\section{Reviewing hyperscaling violating Lifshitz and the 
	shear diffusion constant}

Here we review the discussion in \cite{Kolekar:2016pnr}. 
We are considering nonrelativistic holographic backgrounds described by 
a $(d+1)$-dim hyperscaling violating metric at finite temperature,
\be\label{hvmetric}
ds^2 =  r^{2\theta/d_i}\Big(-\frac{f(r)}{r^{2z}}dt^2 + 
\frac{dr^2}{r^2f(r)}+\frac{\sum_{i=1}^{d_i} dx_{i}^{2}}{r^2}\Big) ,
\qquad  d_i=d-1 ,\qquad d_{eff}=d_i-\theta\ ,
\ee
where $f(r) = 1-(r_0r)^{d+z-\theta-1}$\ and $z$ is the Lifshitz dynamical 
exponent with $\theta$ the hyperscaling violation exponent. The 
temperature of the dual field theory is
\begin{equation}\label{hvtemperature}
T = \frac{(d+z-\theta-1)}{4\pi}~r_0^z\ .
\end{equation}
Here $d_i$ is the boundary spatial dimension while $d_{eff}$ is the 
effective spatial dimension governing various properties of these 
theories, for instance the entropy density $s\sim T^{d_{eff}/z}$.
The null energy conditions following from (\ref{hvmetric}) constrain 
the exponents, giving
\be\label{energyconds}
(d-1-\theta) \big((d-1)(z-1)-\theta\big) \geq 0\ ,\qquad\quad 
(z-1)(d-1+z-\theta) \geq 0\ .
\ee
In \cite{Kovtun:2003wp}, Kovtun, Son and Starinets formulated charge 
and shear diffusion for black brane backgrounds in terms of
long-wavelength limits of perturbations on an appropriately defined
\emph{stretched horizon}, the broad perspective akin to the membrane
paradigm \cite{Thorne:1986iy}. Their quite general analysis begins 
with 
\begin{equation}\label{genmetric}
ds^2 = G_{\mu\nu} dx^\mu dx^\nu = 
G_{tt}(r)dt^2+G_{rr}(r)dr^2+G_{xx}(r)\sum dx_i^2\ ,\qquad\quad 
i=1,\ldots,d_i\ .
\end{equation}
This includes the hyperscaling violating backgrounds \eqref{hvmetric} 
as a subfamily. Charge difffusion of a gauge field perturbation 
$\scriptA_\mu$ in the background \eqref{genmetric} is encoded by the 
charge diffusion constant ${\cal D}$, defined through Fick's Law 
$j^{i}=-\mathcal{D}\partial_{i}j^{t}$, 
where the 4-current $j^{\mu}$ is defined on the stretched horizon 
$r=r_h$ (with $n$ the normal) as $j^{\mu}= n_{\nu}\scriptF^{\mu \nu}|_{r=r_h}$.
Then current conservation $\del_\mu j^\mu = 0$ leads to the diffusion 
equation $\del_tj^t = -\del_ij^i = \mathcal{D} \del_i^2 j^t$, with 
$\mathcal{D}$ the corresponding diffusion constant. Fick's law in 
turn can be shown to apply if the stretched horizon is localized
appropriately with regard to the parameters $\Gamma, q, T$.
Translation invariance along $x\in \{x_i\}$ allows considering plane
wave modes for the perturbations\ $\propto e^{-\Gamma t + iqx}$, where
$\Gamma$ is the typical time scale of variation and $q$ the
$x$-momentum. In the IR regime, the modes vary slowly: this
hydrodynamic regime is a low frequency, long wavelength regime.  The
diffusion of shear gravitational modes can be mapped to charge
diffusion \cite{Kovtun:2003wp}: under Kaluza-Klein compactification of
one of the directions along which there is translation invariance,
tensor perturbations in the original background map to vector
perturbations on the compactified background.
A similar analysis adapting \cite{Kovtun:2003wp} was carried out 
in \cite{Kolekar:2016pnr} for the shear diffusion constant in the 
backgrounds (\ref{hvmetric}), obtaining an effective diffusion equation 
for the metric fluctuations $h_{xy}$ and $h_{ty}$ ($x\equiv x_1$,\
$y\equiv x_2$) around \eqref{genmetric}, depending only on $t, r, x$, 
\ie\ $h_{ty}=h_{ty}(t,x,r) , \ h_{xy}=h_{xy}(t,r,x)$.\
$y$-translation invariance allows a $y$-compactification: then the 
modes $h_{xy}$ and $h_{ty}$ become components of a $U(1)$ gauge field 
in the dimensionally reduced $d$-dim spacetime, with
\be\label{yCompax}
{g}_{\mu \nu} = {G}_{\mu \nu}(G_{xx})^{\frac{1}{d-2}} \qquad 
[\mu, \nu = 0,\ldots, d-1];\qquad
\scriptA_t = (G_{xx})^{-1}h_{ty}\ ,\quad \scriptA_x = (G_{xx})^{-1}h_{xy}\ ,
\ee
where $G_{\mu \nu}$ is the metric given by \eqref{genmetric}. The 
compactified gravitational action contains the Maxwell action,\
$\sqrt{-G}{R}\ \ra\ -\frac{1}{4}\sqrt{-g}\scriptF_{\alpha \beta}\scriptF_{\gamma \delta}
{g}^{\alpha \gamma}{g}^{\beta \delta}(G_{xx})^{\frac{d-1}{d-2}}$,\ with an 
$r$-dependent coupling constant.\
The gauge field equations following from the action are
\be\label{gaugefieldaction}
\partial_{\mu}\Big(\frac{1}{\geff^2}\sqrt{-g}\scriptF^{\mu \nu}\Big)=0\ ,
\qquad\quad {1\over \geff^2} = G_{xx}^{\frac{d-1}{d-2}}\ ,
\ee
where we have read off the $r$-dependent $\geff$ from the compactified 
action. Analysing these Maxwell equations and the Bianchi identity 
assuming gauge field ansatze\ $\scriptA_{\mu}=\scriptA_{\mu}(r)e^{-\Gamma t+iqx}$ and radial gauge $\scriptA_r=0$ as in \cite{Kovtun:2003wp} shows interesting 
simplifications in the near-horizon region.
When $q=0$, these lead to\ 
$\partial_r \big(\frac{\sqrt{-g}}{\geff^2}g^{rr}g^{tt}\partial_r 
\scriptA_t\big)=0$.
We impose the boundary condition that the gauge fields vanish at 
$r=r_c\sim 0$.
As in \cite{Kovtun:2003wp}, for $q$ nonzero but small, we assume an 
ansatz  $\scriptA_t = \scriptA_t^{(0)}+ \scriptA_t^{(1)}+ \ldots,\ 
\scriptA_t^{(1)} = O({q^2\over T^{2/z}})$ as a series expansion in 
$\frac{q^2}{T^{2/z}}$, and likewise for $\scriptA_x$.
The $\scriptA_x$ solution is then found by using the $\scriptA_t^{(0)}$ 
solution and one of the Maxwell equations in the compactified theory. 
We further impose a second assumption
\begin{equation}\label{ansatz2}
|\partial_t \scriptA_x| \ll |\partial_x \scriptA_t|
\end{equation}
as in \cite{Kovtun:2003wp}. For generic values 
\be\label{genExp}
d-z-\theta > -1\ ,
\ee
the leading solution for $\scriptA_t^{(0)}$ has power law behaviour
\be\label{At0soln2}
\scriptA_t^{(0)} = \frac{C}{d-z-\theta+1} e^{-\Gamma t+iqx} 
\ r^{d-z-\theta+1}\ ,
\ee
and
\be\label{Ax0soln}
\scriptA_x^{(0)} = -\frac{i\Gamma}{q}Ce^{-\Gamma t+iqx}
\frac{r_0^{\theta+1-d-z}}{\theta+1-d-z}
\log\left(1-(r_0r)^{d+z-\theta-1}\right)\ .
\ee
As discussed in \cite{Kolekar:2016pnr}, self-consistency of these 
equations and the series solutions holds in the regime
\begin{equation}\label{combinedcriteria}
e^{-\frac{T^{2/z}}{q^2}} \ll \frac{\frac{1}{r_0}-r_h}{\frac{1}{r_0}} \ll 
\frac{q^2}{T^{2/z}} \ll 1\ ,
\end{equation}
for the stretched horizon $r_h$, and the parameters $q, \Gamma$ and 
$T$ (equivalently $r_0$).\ This enables us to define Fick's law on 
the stretched horizon, and thereby the diffusion equation. 
The shear diffusion constant then becomes
\begin{equation}
\label{chargediff}
{\cal D} =\frac{\sqrt{-g(r_h)}}{\geff^{2}(r_h)g_{xx}(r_h)
	\sqrt{-g_{tt}(r_h)g_{rr}(r_h)}}\int_{r_c}^{r_h} dr 
\frac{-g_{tt}(r)g_{rr}(r)\geff^{2}(r)}{\sqrt{-g(r)}}\ ,
\end{equation}
where $r_c$ is the location of the boundary, and we are evaluating 
${\cal D}$ at the stretched horizon.
For a hyperscaling violating theory with $d-z-\theta>-1$, we obtain  
\begin{equation}\label{Dhvgen}
\mathcal{D}\ =\ {1\over r_h^{d-\theta-1}}\int_{r_c}^{r_h} r^{d-z-\theta} dr\ 
= \ {r_h^{2-z}\over d-z-\theta+1}\ \simeq\ 
{r_0^{z-2}\over d-z-\theta+1}\ +\ O(q^2)\ ,
\end{equation}
where we have dropped the contribution in the integral from $r_c$ since 
the UV scale $r_c\ll r_h$ is well-separated from the horizon 
scale. The diffusion constant in (\ref{chargediff}), (\ref{Dhvgen}), 
is evaluated at the stretched horizon $r_h$: however 
$r_h\sim {1\over r_0} + O(q^2)$ so that to leading order ${\cal D}$ is
evaluated at the horizon ${1\over r_0}$. 

In the present hyperscaling violating case, we have seen that\ 
$T \sim r_0^{z}$\ and\ $\mathcal{D} \sim r_0^{z-2}$\ so the product 
$\mathcal{D}T \sim r_0^{2(z-1)}$\ is not dimensionless. 
Using (\ref{hvtemperature}), we have
\be\label{DT(2-z)/z}
{\cal D} = {1\over d-z-\theta+1} 
\Big({4\pi\over d+z-\theta-1}\Big)^{{z-2\over z}}\ 
T^{{z-2\over z}}\ ,
\ee
as the scaling with temperature $T$ of the leading diffusion constant 
(\ref{Dhvgen}). See also \eg\ \cite{Pang:2009wa,Cremonini:2011ej,
	Sadeghi:2014zia,Roychowdhury:2014lta,Kuang:2015mlf,Blake:2016wvh} for 
previous investigations including via holography.

This motivates us to guess the universal relation
\be\label{eta/s,DT(2-z)/zHV}
{\eta\over s}\ =\ 
{(d-z-\theta+1)\over 4\pi} \mathcal{D} r_0^{2-z}\ =\ 
{(d-z-\theta+1)\over 4\pi} \Big({4\pi\over d+z-\theta-1}\Big)^{{2-z\over z}}\ 
\mathcal{D} T^{{2-z\over z}}\ =\ {1\over 4\pi}
\ee
between $\eta, s, {\cal D}, T$, for general exponents $z,\theta$.\ As 
discussed in \cite{Kolekar:2016pnr}, this is consistent with 
relativistic theories ($\theta=0, z=1$) arising from $AdS$ and with 
theories with exact Lifshitz scaling symmetry,\ 
$x_i\ra \lambda x_i,\ t\ra \lambda^z t$.
Then the diffusion equation $\del_t j^t = {\cal D} \del_i^2 j^t$ shows the 
diffusion constant to have scaling dimension $dim[{\cal D}]=z-2$,\ where 
momentum scaling is $[\del_i]=1$\ (or equivalently, $[x_i]=-1,\ [t]=-z$).
With temperature scaling as inverse time, we have $dim[T]=z$. For 
hyperscaling violating theories with $z=1$, it can be seen that\
${\cal D} = {1\over 4\pi T}$, with the $\theta$-dependent prefactors
cancelling precisely. Thus all hyperscaling violating theories with
$z=1$ appear to satisfy the universal viscosity bound\
${\eta\over s} = {\cal D} T = {1\over 4\pi}$~.

When $d-z-\theta=-1$, we obtain logarithmic behaviour
\be\label{violation}
\scriptA_t^{(0)} = C e^{-\Gamma t+iqx}\ \log\Big({r\over r_c}\Big)\quad
\longrightarrow\quad 
\mathcal{D} = r_0^{d-\theta-1}\ \log\Big(\frac{1}{r_0r_c}\Big) = 
r_0^{z-2} \ \log\Big(\frac{1}{r_0r_c}\Big)\ .
\ee
This implies that in the low temperature limit $r_0\ra 0$, the diffusion 
constant becomes vanishingly small if $d_i-\theta>0$, or equivalently 
$z>2$.\ However further analysis as in \cite{Kolekar:2016pnr} reveals
that the near-horizon expansion is less reliable in this case, 
necessitating more investigation.

\section{Perturbations in the absence of gauge field: Dilaton gravity}

In this section, we will analyse the perturbations in hyperscaling 
violating Lifshitz theories focussing on 4 bulk dimensions (\ie\ $d=3,\ 
d_i=2$) for simplicity and concreteness. The hyperscaling violating 
metric is
\begin{equation}
ds^2= r^{\theta}\left(-\frac{f(r)}{r^{2z}}dt^2+\frac{dr^2}{f(r)r^2}+\frac{dx^2+dy^2}{r^2} \right)\ , \quad \qquad d_i=2\ , \qquad d_{eff}=2-\theta\ ,
\end{equation}
where $f(r)=1-(r_0r)^{2+z-\theta}$. The temperature for the dual field
theory (\ie\ the Hawking temperature for the black brane) is
$T=\frac{2+z-\theta}{4\pi}r_0^z$. We will make a gauge choice for the
perturbations by setting $h_{\mu r}=0$ (radial gauge) and assume that
the perturbations to be of the form $h_{\mu \nu}(t,x,r) = e^{-i \omega
  t+iq\cdot x}h_{\mu \nu}(r)$ where $x$ is one of the spatial
directions in the boundary theory. The shear mode $h_{xy}$ couples to
$h_{ty}$ and decouples from the scalar mode $\varphi$ giving us a
system of three coupled equations,
\begin{equation}
\label{dg-upstairseom1}
\partial_r(r^{z+\theta-3}\partial_r(r^{2-\theta}h_{ty}))-\frac{r^{z+\theta-3}}{f}q(\omega r^{2-\theta}h_{xy}+qr^{2-\theta}h_{ty})=0\ ,
\end{equation}
\begin{equation}
\label{dg-upstairseom2}
\partial_r(r^{-1-z+\theta}f\partial_r(r^{2-\theta}h_{xy}))+\frac{r^{z+\theta-3}}{f}\omega(\omega r^{2-\theta}h_{xy}+qr^{2-\theta}h_{ty})=0\ ,
\end{equation}
\begin{equation}
\label{dg-upstairseom3}
q\partial_r(r^{2-\theta}h_{xy})+\frac{\omega}{f}r^{2z-2}\partial_r(r^{2-\theta}h_{ty})=0\ .
\end{equation}
In terms of the $y$-compactified theory variables 
\be\label{dg-up-down-map}
\begin{aligned}
	&{g}_{\mu \nu} = r^{\theta-2}{G}_{\mu \nu} \qquad 
	[\mu, \nu = t,x,r];\qquad
	\scriptA_t = r^{2-\theta}h_{ty}\ ,\quad \scriptA_x = r^{2-\theta}h_{xy}\ ,\\
	&\scriptF_{rt} = \partial_r(r^{2-\theta}\scriptA_t)\ , \qquad \quad \scriptF_{rx}=\partial_r(r^{2-\theta}\scriptA_x)\ , \qquad \quad \scriptF_{tx}=-ir^{2-\theta}(\omega h_{xy}+qh_{ty})\ ,
\end{aligned}
\ee
the above linearized Einstein equations become
\begin{eqnarray}
\label{dg-nu-t-eqn}
\sqrt{-g}e^{4\psi}g^{tt}g^{xx}\partial_{x}\scriptF_{tx}+\partial_r(\sqrt{-g}e^{4\psi}g^{rr}g^{tt}\scriptF_{tr})&=&0 ,\\
\label{dg-nu-x-eqn}
\sqrt{-g}e^{4\psi}g^{tt}g^{xx}\partial_{t}\scriptF_{tx}+\partial_r(\sqrt{-g}e^{4\psi}g^{rr}g^{xx}\scriptF_{rx})&=&0\ ,\\
g^{tt}\partial_t
\label{dg-nu-r-eqn}
\scriptF_{tr}+g^{xx}\partial_{x}\scriptF_{xr}&=&0\ ,
\end{eqnarray}
where $e^{4\psi}=\frac{1}{\geff^2}=r^{2\theta-4}$. Other than these, we also have a Bianchi Identity
\begin{equation}\label{dg-bianchi}
\partial_t\scriptF_{rx}+\partial_x\scriptF_{tr}-\partial_r{\scriptF}_{tx}=0\ ,
\end{equation}
which is a trivial relation in the higher dimensional theory.
Equation \eqref{dg-upstairseom3} is a constraint equation in the
higher dimensional theory which can be mapped to \eqref{dg-nu-r-eqn}
in the $y$-compactified theory.  Defining currents as
$j^{\nu}=n_{\mu}\scriptF^{\mu \nu}$ ($n_{\mu}$ being the normal vector
to the boundary $r=r_c$, with $g^{rr}n_r^2=1$) we can write them in 
terms of the perturbations of the higher dimensional theory,
\begin{eqnarray}
\label{horizoncurrent1}
j^{x}&=&n_r\scriptF^{xr}=r^{6-3\theta}\sqrt{f}\ \partial_r(r^{2-\theta}h_{xy})\ ,\\
\label{horizoncurrent2}
j^{t}&=&n_r\scriptF^{tr}=-\frac{r^{4+2z-3\theta}}{\sqrt{f}}\ \partial_r(r^{2-\theta}h_{ty})\ .
\end{eqnarray}
Identifying the ratio $\mathcal{D} \equiv -\frac{\omega}{iq^2}$ we can 
essentially write \eqref{dg-upstairseom3} in the form of Fick's Law as
\begin{equation}
j^x = -\mathcal{D}\partial_x j^{t}\ .
\end{equation}

The formulation of Fick's Law in \cite{Kolekar:2016pnr,Kovtun:2003wp} 
is done entirely in terms of field variables of the $y$-compactified theory. 
Differentiating \eqref{dg-nu-r-eqn} w.r.t $t$ we can eliminate 
$\scriptF_{rx}$ using the Bianchi Identity \eqref{dg-bianchi} to get 
the following equation
\begin{equation}\label{dg-ME1}
\partial_t^2{\scriptF}_{tr}+r^{2-2z}f\partial_x(-\partial_x{\scriptF}_{tr}+\partial_r{\scriptF}_{tx})=0\ .
\end{equation}
In the the near horizon region approximating the thermal factor as 
$f(r) \approx (2+z-\theta)\frac{(1/r_0)-r}{1/r_0}$ and parametrizing 
the frequency as $\omega = -i\Gamma$ for some positive $\Gamma$ so that 
the perturbations decay in time, \eqref{dg-ME1} can be written as
\begin{equation}\label{dg-ftr-eqn}
\left(1+(2+z-\theta) r_0^{2z-2}\frac{q^2}{\Gamma^2}\cdot \frac{\frac{1}{r_0}-r}{\frac{1}{r_0}}\right){\scriptF}_{tr}\approx-(2+z-\theta)r_0^{2z-2}\frac{iq}{\Gamma^2}\cdot\frac{\frac{1}{r_0}-r}{\frac{1}{r_0}}\partial_r{\scriptF}_{tx}\ .
\end{equation} 
Assuming
\begin{equation}\label{outerbound}
\frac{\frac{1}{r_0}-r}{\frac{1}{r_0}}\ll \frac{\Gamma^2}{q^2r_0^{2z-2}}\ ,
\end{equation}
we differentiate both sides w.r.t $x$ and approximate \eqref{dg-ftr-eqn} 
further as
\begin{equation}\label{dg-ftrestimate}
\partial_x {\scriptF}_{tr}\approx (2+z-\theta)\frac{q^2r_0^{2z-2}}{\Gamma^2}\cdot \frac{\frac{1}{r_0}-r}{\frac{1}{r_0}}\partial_r {\scriptF}_{tx}\ .
\end{equation}
The assumption \eqref{outerbound} implies
\begin{equation}\label{fieldstrcomparison}
\partial_x {\scriptF}_{tr}\ll \partial_r {\scriptF}_{tx}\ ,
\end{equation}
which in turn simplifies the Bianchi Identity (\ref{dg-bianchi}) to
\begin{equation}\label{dg-approx-bianchi}
\partial_t \scriptF_{rx}= \partial_x{\scriptF}_{rt}+\partial_r{\scriptF}_{tx}\
\sim\ \partial_r {\scriptF}_{tx}\ .
\end{equation}

Differentiating \eqref{dg-nu-x-eqn} w.r.t $t$ 
we get
\begin{equation}
\partial_r(r^{\theta-z-1}f \partial_t\scriptF_{rx})-\frac{r^{z+\theta-3}}{f}\partial_t^2 {\scriptF}_{tx}=0\ .
\end{equation}
Using the approximate Bianchi identity \eqref{dg-approx-bianchi}, to 
substitute for $\scriptF_{rx}$ and then multiplying throughout with 
$-\frac{f}{r^{z+\theta-3}}$ we obtain a wave equation for the field strength 
$\scriptF_{tx}$
\begin{equation}
\label{waveequation-ftx}
\partial_t^2 {\scriptF}_{tx}-\nu^2\left(\frac{1}{r_0}-r\right)\partial_r \left(\left(\frac{1}{r_0}-r\right)\partial_r {\scriptF}_{tx} \right)\approx 0\ ,
\end{equation}
where $\nu$ is given by
\begin{equation}\label{nu}
\nu=(2+z-\theta)r_0^z\ .
\end{equation}
The horizon is a one-way membrane: we incorporate this by requiring that
all perturbations obey ingoing boundary conditions at the horizon. This 
dissipative feature is of course at the heart of the diffusion equation 
that results from this near-horizon perturbations analysis. Thus,
imposing ingoing boundary conditions on the wave equation amounts to 
choosing the ingoing solution, leading to
 \begin{equation}\label{ingoingSolnFtx}
 \scriptF_{tx} = 
f_1\left(t+\frac{1}{\nu}\log\left(\frac{1}{r_0}-r\right)\right)\ ,
 \end{equation}
where $f_1$ is any arbitrary smooth function. If we now ensure that 
the perturbations decay as $t \rightarrow \infty$ we obtain
 \begin{equation}\label{dg-ftx-frx-reln}
 \scriptF_{tx}+\nu\Big(\frac{1}{r_0}-r\Big)\scriptF_{rx}=0\ .
\end{equation}
As reviewed in sec.~2, the leading solutions for $\scriptA_t, \scriptA_x$ are
\bea\label{ansatz1}
\scriptA_t^{(0)} = Ce^{-\Gamma t+iqx}\int_{r_c}^{r}dr' 
\frac{g_{tt}(r')g_{rr}(r')}{\sqrt{-g(r')}}\cdot \geff^2(r')\ 
= Ce^{-\Gamma t+iqx}\int_{r_c}^{r}dr'\ 
\frac{G_{tt}(r')G_{rr}(r')}{G_{xx}(r')\sqrt{-G(r')}}\ ,\nonumber\\
\scriptA_x^{(0)} = -\frac{i\Gamma}{q}Ce^{-\Gamma t+iqx}\int_{r_c}^{r}dr' 
\frac{g_{xx}(r')g_{rr}(r')}{\sqrt{-g(r')}}\cdot \geff^{2}(r') 
= -\frac{i\Gamma}{q}Ce^{-\Gamma t+iqx}\int_{r_c}^{r}dr' 
\frac{G_{rr}(r')}{\sqrt{-G(r')}}\ ,\ \
\eea
where $r_c\sim 0$ is the boundary where we impose the boundary 
conditions that the perturbations die.
Above, we have used (\ref{yCompax}), (\ref{gaugefieldaction}), with 
$C$ some constant: these give the solutions (\ref{At0soln2}) and 
(\ref{Ax0soln}) when (\ref{genExp}) holds. Then from Fick's Law 
using \eqref{dg-ftx-frx-reln} on the stretched horizon we can calculate 
the shear diffusion constant as
\begin{align}\label{diffconst}
\mathcal{D}\equiv -\frac{j^x}{\partial_x j^t}&=-\frac{g_{tt}}{g_{xx}}\frac{\scriptF_{rx}}{\partial_x\scriptF_{rt}} \approx -r_0^{z-1}\frac{\scriptF_{tx}}{\partial_x\scriptF_{rt}}  =r_0^{z-1}\left.\frac{\scriptA_t}{\scriptF_{rt}}\right|_{r\sim r_h}\simeq \ 
{r_0^{z-2}\over 4-z-\theta}\ \ .
\end{align}

As should be clear, a key ingredient that goes in the formulation of
Fick's Law is the relation \eqref{dg-ftx-frx-reln} for the field
strengths $\scriptF_{tx}$ and $\scriptF_{rx}$. In the context of the
higher-dimensional hyperscaling violating theory where the perturbations 
satisfy (\ref{dg-upstairseom1}), (\ref{dg-upstairseom2}), 
(\ref{dg-upstairseom3}), this relation can be derived exactly without 
any assumptions on the parameters $q$ and $\omega$. It turns out in 
this context it is a consequence of imposing a certain physical 
condition on the function $H(t,r,x)$ defined as
\begin{equation}
H(t,r,x) \equiv r^{\theta-z-1}f \cdot \partial_r (r^{2-\theta}h_{xy})\ .
\end{equation}
This condition that we impose is given by
\begin{equation}
\label{infallH}
(\partial_t+ f \cdot r^{1-z}\partial_r)\ H=0\ .
\end{equation}
Defining two new coordinates $u$ and $v$ as
\begin{eqnarray}
v&=&t+ \frac{1}{\nu}\log \left(\frac{1}{r_0}-r\right)\ , \nonumber\\
u&=&t- \frac{1}{\nu}\log \left(\frac{1}{r_0}-r\right)\ .
\end{eqnarray}
For $r\ll {1\over r_0}$ expanding the log, we see that 
$v\sim t-{r_0\over\nu}r$ so $v$ is the ingoing coordinate (with $r$ 
increasing towards the interior).
We see that in the near horizon region the full wave operator is 
\begin{equation}
4\partial_u \partial_v \equiv \partial_t^2 -\nu^2\left(\frac{1}{r_0}-r\right)\partial_r \left(\left(\frac{1}{r_0}-r\right)\partial_r \right)\ ,
\end{equation}
while the linear differential operator acting on $H$ in
\eqref{infallH} is essentially $\partial_t+ f \cdot r^{1-z}\partial_r
\approx \partial_t+\nu \left(\frac{1}{r_0}-r\right)\partial_r
= \partial_u$ . With $v$ the ingoing coordinate, this can be thus 
interpreted as the ingoing condition $\del_uH=0$ implying that the 
function has the form $H=H(v)$.\\
Likewise, choosing the solution (\ref{ingoingSolnFtx}) is equivalent to 
requiring that the field strength $\scriptF_{tx}$ obeys the ingoing 
condition
\begin{equation}\label{ftx-ingoing}
\partial_t {\scriptF}_{tx}+\nu\left(\frac{1}{r_0}-r\right)\partial_r {\scriptF}_{tx}=0\ ,
\end{equation}
which can also be written as\ 
$\partial_u \scriptF_{tx}=0$, giving\ $\scriptF_{tx}=\scriptF_{tx}(v)$. 
Using \eqref{dg-up-down-map} we can write
\begin{equation}\label{fldstrngths}
\partial_t \scriptF_{tx}=- r^{2-\theta}\omega(\omega h_{xy}+q h_{ty}), \ \qquad 
\scriptF_{rx}=\partial_r (r^{2-\theta}h_{xy})=\frac{r^{z+1-\theta}}{f}H\ .
\end{equation}
The above equalities in conjunction with \eqref{dg-upstairseom2}
gives
\begin{equation}
\label{ftx-H-reln}
\begin{aligned}
\partial_r H= -\frac{r^{z-1}}{f}\omega(\omega h_{xy}+qh_{ty})=\frac{r^{z+\theta-3}}{f}\partial_t \scriptF_{tx}\ .
\end{aligned}
\end{equation}
Also \eqref{infallH} naturally implies
\be
\partial_rH=-\frac{r^{z-1}}{f}\partial_t H=-r^{\theta-2}\partial_t
\scriptF_{rx}\ .
\ee
Equating the above two expressions for $\partial_r H$, we recover the
relation \eqref{dg-ftx-frx-reln} as was obtained in
\cite{Kolekar:2016pnr}. It should be noted that the relation between
$\scriptF_{rx}$ and $\scriptF_{tx}$ was obtained in the
$y$-compactified theory by making certain self-consistent
approximations involving the parameters $q$ and $\omega$ which is
quite distinct from the derivation demonstrated here, using 
(\ref{dg-upstairseom1}), (\ref{dg-upstairseom2}), 
(\ref{dg-upstairseom3}), directly.

The leading order value of the diffusion constant $\mathcal{D}$ is 
given by $\scriptA_t^{(0)}$, which is obtained by solving the $q=0$ and 
$\omega=0$ sector of \eqref{dg-upstairseom1} 
\begin{equation}
\partial_r(r^{z+\theta-3}\partial_r(r^{2-\theta}h_{ty}))=0\ .
\end{equation}
The solution to the above equation is given by
\begin{equation}
h_{ty}(r)=c_{1}r^{\theta-2}+c_{2}r^{2-z}\ ,
\end{equation}
where $c_1$ and $c_2$ are arbitrary constants. For $z< 4-\theta$, 
the $\theta-2$ fall-off (non-normalizable mode) dominates over the 
$2-z$ fall-off (normalizable mode) near the boundary $r\sim r_c$, 
while for $z> 4-\theta$ we see the exact opposite behaviour. When 
$z=4-\theta$ there is a degeneracy in the two fall-offs and we have 
a new independent solution which scales logarithmically with $r$,
\begin{equation}
h_{ty}(r)=c_1r^{\theta-2}+c_{2}r^{\theta-2}\log {r\over r_c}\ .
\end{equation}
The swapping of roles between the normalizable and non-normalizable 
modes around the point $z=4-\theta$ gives some insight into the 
unusual logarithmic scaling for the diffusion constant when $z=4-\theta$. 
It is in fact reminiscent of the alternative quantization of 
field modes \cite{Klebanov:1999tb} and thus holographically it is 
not surprising that the relevant correlation function exhibits 
logarithmic behaviour.

In the presence of a background gauge field, the analysis changes
significantly. The analog of \eqref{dg-upstairseom3} including the
gauge field is given by
\begin{equation}
q\partial_r(r^{2-\theta}h_{xy})+\frac{\omega}{f}r^{2z-2}\partial_r(r^{2-\theta}h_{ty})-k\frac{\omega}{f}r^{z-\theta+1}a_y=0\ .
\end{equation}
Due to the presence of the gauge field perturbation $a_y$, we cannot
map the above equation to Fick's Law by defining horizon currents as
before.  Subsequently we will show that a field redefinition which
involves a non-trivial combination of $h_{ty}$ and $\int a_y\ dr$
gives us an equation which is similar in structure to Fick's Law in
the dimensionally reduced theory.

\section{Perturbations to hyperscaling violating spacetime}

We are considering nonrelativistic holographic backgrounds described
by a $(d+1)$-dimn hypercsaling violating metric at finite temperature
as given in \eqref{hvmetric}. The metric \eqref{hvmetric} is a
solution to the action
\begin{equation}
\label{hvaction}
S= -\frac{1}{16 \pi G_{N}^{(d+1)}}\int d^{d+1}x \sqrt{-G}\left[R-\frac{1}{2}\partial_{\mu}\phi \partial^{\mu}\phi-\frac{Z(\phi)}{4}F_{\mu \nu}F^{\mu \nu}+V(\phi) \right]\ ,
\end{equation}
where $\phi$ is the dilaton with a potential 
$V(\phi)=-2\Lambda e^{-\delta \phi}$, where
\begin{equation}
\delta=\dfrac{2\theta/d_i}{\sqrt{2(d_i-\theta)(z-\theta/d_i-1)}} \quad \mbox{and} \quad \Lambda=-\frac{1}{2}(d_i+z-\theta)(d_i+z-\theta -1)\ . 
\end{equation} 
The background gauge field is given by
\begin{equation}
A_t = \frac{\alpha f(r)}{r^{d_i+z-\theta}}\ , \qquad \alpha=-\sqrt{\frac{2(z-1)}{d_i+z-\theta}}
\end{equation}
and with gauge field coupling being
\begin{equation} 
Z(\phi)=e^{\lambda \phi}=r^{\frac{2\theta}{d_i}+2d_i-2\theta}\ , \qquad \mbox{where} \qquad \lambda=\dfrac{2\theta/d_i+2d_i-2\theta}{\sqrt{2(d_i-\theta)(z-\theta/d_i-1)}}\ .
\end{equation}
Since only the $A_{t}(r)$ component is non-zero, we have only one
non-zero field strength
\begin{equation}
F_{rt}=\frac{-\alpha (d_{i}+z-\theta)}{r^{d_{i}+z-\theta+1}}\ .
\end{equation}
Varying the action \eqref{hvaction} with respect to the bulk metric
$G_{\mu \nu}$, the gauge field $A_{\mu}$ and $\phi$ we get the
following equations of motion
\begin{equation}\label{einsteineqn}
R_{\mu\nu}=\frac{1}{2}\partial_{\mu}\phi\partial_{\nu}\phi -G_{\mu\nu}\frac{V(\phi)}{d-1} + \frac{Z(\phi)}{2}G^{\rho\sigma}F_{\rho\mu}F_{\sigma\nu} - \frac{Z(\phi)}{4(d-1)}G_{\mu\nu}F_{\rho\sigma}F^{\rho\sigma}\ ,
\end{equation}
\begin{equation}\label{gaugeeqn}
\nabla_{\mu}(Z(\phi)F^{\mu\nu})=0\ ,
\end{equation}
\begin{equation}
\frac{1}{\sqrt{-G}}\partial_{\mu}(\sqrt{-G}G^{\mu\nu}\partial_{\nu}\phi)+\frac{\partial V(\phi)}{\partial\phi}-\frac{1}{4}\frac{\partial Z(\phi)}{\partial\phi}F_{\rho\sigma}F^{\rho\sigma}=0\ .
\end{equation}

We turn on generalized gravitational, gauge field and scalar field
perturbations $h_{\mu \nu}(\bar{x},r)$, $a_{\mu}(\bar{x},r)$ and
$\varphi(\bar{x},r)$ where $\bar{x}$ denotes all the boundary
coordinates collectively. Later, we will make a certain gauge choice
(radial gauge) for the perturbations in order to simplify our
calculations.  At the linearized level, the Einstein's equations
\eqref{einsteineqn} are given by
\begin{equation}
\begin{split}\label{lineinsteineqn}
R^{(1)}_{\mu\nu}=&\frac{1}{2}\partial_{\mu}\phi\partial_{\nu}\varphi+\frac{1}{2}\partial_{\mu}\varphi\partial_{\nu}\phi-\frac{V}{2}(h_{\mu\nu}- G_{\mu\nu}\delta\varphi)\\
&+\frac{Z}{2}\left[G^{\rho\sigma}F_{\mu\rho}f_{\nu\sigma}+G^{\rho\sigma}f_{\mu\rho}F_{\nu\sigma}-h^{\rho\sigma}F_{\mu\rho}F_{\nu\sigma}+\lambda\varphi G^{\rho\sigma}F_{\mu\rho}F_{\nu\sigma} \right]\\
&-Z\left[\frac{1}{4}G_{\mu\nu}(F_{\rho\sigma}f^{\rho\sigma}-g^{\rho\alpha}h^{\sigma\beta}F_{\rho\sigma}F_{\alpha\beta})+\frac{1}{8}h_{\mu\nu}F_{\rho\sigma}F^{\rho\sigma}+\frac{1}{8}\lambda\varphi G_{\mu\nu}F_{\rho\sigma}F^{\rho\sigma} \right]\ ,
\end{split}
\end{equation}
where
\begin{equation}
\label{lingravity}
R^{(1)}_{\mu \nu}=\frac{1}{2}[\nabla_{\alpha}\nabla_{\nu}h^{\alpha}_{\mu}+\nabla_{\alpha}\nabla_{\mu}h^{\alpha}_{\nu}-\nabla_{\alpha}\nabla^{\alpha}h_{\mu \nu}-\nabla_{\nu}\nabla_{\mu}h]; \quad f_{\mu \nu}= \partial_{\mu}a_{\nu}-\partial_{\nu}a_{\mu}; \quad h=G^{\mu\nu}h_{\mu\nu}\ .
\end{equation}
Similarly, the Maxwell's Equations \eqref{gaugeeqn} upto linearized
order gives the following equations of motion
\begin{equation}\label{lingaugeeqn}
\nabla_{\mu}(Z\,f^{\mu \nu})-\nabla_{\mu}(Z\,h^{\mu \rho}F_{\rho}^{\ \ \nu})-Z(\nabla_{\mu}h^{\nu \sigma})F^{\mu}_{\ \ \sigma} +\frac{1}{2}(\nabla_{\mu}h)Z\,F^{\mu \nu}+\lambda\, Z\,F^{\mu \nu}\partial_{\mu} \varphi=0\ .
\end{equation}
Finally, the linearized scalar field equation is:
\begin{equation}
\begin{split}\label{linscalareqn}
\frac{1}{\sqrt{-G}}\partial_{\mu}(\sqrt{-G}G^{\mu \nu}\partial_{\nu}\varphi) &-\frac{1}{\sqrt{-G}}\partial_{\mu}(\sqrt{-G}h^{\mu \nu}\partial_{\nu}\phi)+\frac{1}{2}G^{\mu \nu}\partial_{\nu}\phi\partial_{\mu}h  +V\delta^{2}\varphi\\
&-\frac{\lambda Z}{4}(2F_{\mu\nu}f^{\mu\nu}-2G^{\mu\rho}h^{\nu\sigma}F_{\mu\nu}F_{\rho\sigma}+\lambda\varphi F_{\mu\nu}F^{\mu\nu})=0\ .
\end{split}
\end{equation}
In the linearized equations of motion \eqref{lineinsteineqn},
\eqref{lingaugeeqn} and \eqref{linscalareqn}, all indices are raised
with respect to the background metric \eqref{hvmetric}. For the sake
of simplicity our subsequent analysis will be for $d=3$ (\ie\ $d_i=2$)
but we expect this procedure can be generalized for higher dimensions.

\subsection{Perturbations to hyperscaling violating spacetime: 
Einstein-Maxwell-dilaton (EMD) theory in 4 dimensions ($d=3$)}

In the presence of a background gauge field, the perturbations in the
metric sector $h_{xy}$ and $h_{ty}$ couples to perturbation to the
background gauge field $a_y$. For the sake of completeness, we have
also listed the equations of motion for the other perturbations in
\ref{appendix-A}. In the radial gauge (\ie\ $h_{\mu r}=0$) assuming
perturbations of the form $h_{\mu \nu}= e^{-i\omega t+iqx}h_{\mu\nu}(r)$, 
the coupled set of equations governing $h_{ty}$, $h_{xy}$ and $a_y$ 
become
\begin{eqnarray}
\label{upstairseom1}
\partial_r(r^{5-z-\theta}f\partial_ra_y)+\frac{\omega^2}{f}r^{3+z-\theta}a_y-q^2r^{5-z-\theta}a_y-k\partial_r(r^{2-\theta}h_{ty})&=&0\ ,\\
\label{upstairseom2}
\partial_r(r^{z+\theta-3}\partial_r(r^{2-\theta}h_{ty}))-\frac{r^{z+\theta-3}}{f}q(\omega r^{2-\theta}h_{xy}+qr^{2-\theta}h_{ty})-k\partial_ra_y&=&0\ ,\\
\label{upstairseom3}
\partial_r(r^{-1-z+\theta}f\partial_r(r^{2-\theta}h_{xy}))+\frac{r^{z+\theta-3}}{f}\omega(\omega r^{2-\theta}h_{xy}+qr^{2-\theta}h_{ty})&=&0\ ,\\
\label{upstairseom4}
q\partial_r(r^{2-\theta}h_{xy})+\frac{\omega}{f}r^{2z-2}\partial_r(r^{2-\theta}h_{ty})-k\frac{\omega}{f}r^{z-\theta+1}a_y&=&0\ ,
\end{eqnarray}
where 
\begin{equation}
k=(2+z-\theta)\alpha\ , \qquad \qquad \alpha=-\sqrt{\frac{2(z-1)}{2+z-\theta}}\ . 
\end{equation}
Note that the last equation \eqref{upstairseom4} is a constraint 
equation in $r$ which we will eventually use to map to Fick's Law.
Now we will further assume that the solutions to the perturbations 
$h_{xy}$, $h_{ty}$ and $a_y$ can be expanded as a series in 
${q^{2}\over T^{2/z}}$ which we schematically write as
\begin{equation}\label{up-seriesansatz}
\begin{pmatrix}
h_{ty}(t,x,r)\\
h_{xy}(t,x,r)\\
a_y(t,x,r)
\end{pmatrix}
\equiv
\begin{pmatrix}
h_{ty}^{(0)}(t,x,r)\\
h_{xy}^{(0)}(t,x,r)\\
a_y^{(0)}(t,x,r)
\end{pmatrix}+ 
\begin{pmatrix}
h_{ty}^{(1)}(t,x,r)\\
h_{xy}^{(1)}(t,x,r)\\
a_y^{(1)}(t,x,r)
\end{pmatrix}+ \cdots\ , \qquad \begin{pmatrix}
h_{ty}^{(1)}(t,x,r)\\
h_{xy}^{(1)}(t,x,r)\\
a_y^{(1)}(t,x,r)
\end{pmatrix}= O\left(\frac{q^2}{T^{2/z}}\right)\ .
\end{equation}
Subsequently we will show that this formalism is indeed consistent
with the proposed series ansatz. Compactifying along $y$, we can write
the lower dimensional field variables in terms of the fields in the 4
dimensional hyperscaling violating theory as
\begin{equation}
\scriptA_t = r^{2-\theta}h_{ty}\ , \qquad
\scriptA_x = r^{2-\theta}h_{xy}\ , \qquad
\chi =a_y\ , \qquad
g_{\mu \nu}=r^{\theta-2}G_{\mu \nu}\ ,
\end{equation}
and 
\begin{equation}
\scriptF_{\mu \nu}= \partial_{\mu}\scriptA_{\nu}-\partial_{\nu}\scriptA_{\mu},
\qquad Z=r^{4-\theta}, \qquad e^{2\psi}=G_{yy}=r^{\theta-2}\ .
\end{equation}
In terms of the fields defined above, \eqref{upstairseom2}, 
\eqref{upstairseom3} and \eqref{upstairseom4} take the form
\begin{eqnarray}
\label{nu-t-eqn}
\sqrt{-g}e^{4\psi}g^{tt}g^{xx}\partial_{x}\scriptF_{tx}+\partial_r(\sqrt{-g}e^{4\psi}g^{rr}g^{tt}\scriptF_{tr})&=&\sqrt{-g}e^{2\psi}ZF^{rt}\partial_r \chi\ ,\\
\label{nu-x-eqn}
\sqrt{-g}e^{4\psi}g^{tt}g^{xx}\partial_{t}\scriptF_{tx}+\partial_r(\sqrt{-g}e^{4\psi}g^{rr}g^{xx}\scriptF_{rx})&=&0\ ,\\
g^{tt}\partial_t
\label{nu-r-eqn}
\scriptF_{tr}+g^{xx}\partial_{x}\scriptF_{xr}&=&e^{-2\psi}g_{rr}ZF^{rt}\partial_t \chi\ .
\end{eqnarray}
The perturbation to the background gauge field $a_y$ becomes an
effective scalar field $\chi$ in the lower dimensional theory whose
equation of motion is given by
\begin{equation}
\label{down-scalareqn}
-\frac{r^{3+z-\theta}}{f}\partial_t^2 \chi +r^{5-z-\theta}\partial_x^{2}\chi +\partial_r(r^{5-z-\theta}f.\partial_r \chi)-k \scriptF_{rt}=0\ .
\end{equation}
The equations \eqref{nu-t-eqn}, \eqref{nu-x-eqn}, \eqref{nu-r-eqn} and
\eqref{down-scalareqn} can be derived by compactifying
\eqref{hvaction} along $y$ and varying the effective lower dimensional
action with respect $\scriptA_{\mu}$ and $\chi$ as detailed in
\ref{dim-redn}. The field strengths also satisfy the Bianchi identity
\begin{equation}
\label{bianchi-1}
\partial_t\scriptF_{rx}+\partial_x \scriptF_{tr}-\partial_r\scriptF_{tx}=0\ ,
\end{equation}
which is a trivial relation in the higher dimensional theory.

Like the earlier case of dilaton gravity with no gauge field (sec.~3), 
we could define the horizon currents as $j^{\nu}=n_{\mu}\scriptF^{\mu \nu}$,
with the explicit expression for these currents in terms of the higher
dimensional theory as in \eqref{horizoncurrent1},
\eqref{horizoncurrent2}.  However, unlike the earlier case where
\eqref{dg-upstairseom3} was mapped to Fick's Law in the
$y$-compactified theory, we do not observe such a structure for
\eqref{upstairseom4}.
In the presence of a background gauge field, the behaviour of the
perturbations $h_{ty}$ and $a_y$ is expected to be different than before
(sec.~3) since even in the $q=\Gamma=0$ sector, they are coupled. The
equations governing them follows from \eqref{upstairseom1} and
\eqref{upstairseom2},
\be\label{up-leadingeom}
\partial_r(r^{5-z-\theta}f\partial_ra_y)-k\partial_r(r^{2-\theta}h_{ty})=0\ ,
\qquad 
 \partial_r(r^{z+\theta-3}\partial_r(r^{2-\theta}h_{ty}))-k\partial_ra_y=0\ .
\ee
From the expression for the diffusion constant for dilaton gravity
\eqref{diffconst} one might expect that even in this case, the
expression for the diffusion constant will require the detailed
solutions to \eqref{up-leadingeom}. This is a system of two
second-order coupled differential equations: eliminating $a_y$
gives a 3rd order differential equation for $h_{ty}$, and likewise
eliminating $h_{ty}$ leads to a 3rd order equation for $a_y$.
Thus we have 3 independent solutions for each of the functions
$h_{ty}$ and $a_y$. These solutions can be found explicitly but 
we relegate discussing them in detail to Appendix \ref{leading-soln}, 
since it turns out interestingly that the diffusion analysis that 
follows does not depend in detail on them.

In this regard, it is important to note that the hyperscaling
violating Lifshitz black branes here are not charged: the gauge field
and scalar here simply serve as sources that support the
nonrelativistic metric as a solution to the gravity theory. Using
intuition from the fluid-gravity correspondence
\cite{Bhattacharyya:2008jc}, the fact that these are uncharged black
branes means that the near-horizon perturbations must effectively be
characterized simply by local temperature and velocity
fluctuations. Charge cannot enter as an extra variable characterizing
the near-horizon region. Thus the structure of the diffusion equation
and the diffusion constant should not be dramatically altered by the
presence of the gauge field, although the gauge field perturbation
$a_y$ is not ``subleading'' to the $h_{ty}$ perturbation in any sense,
from \eqref{up-leadingeom}, and also the linearized Einstein equations
(\ref{upstairseom1})-(\ref{upstairseom4}).

Armed with this intuition, looking closer, we see that we can rearrange 
\eqref{upstairseom4} to write
\begin{equation}
q\partial_r(r^{2-\theta}h_{xy})+\frac{\omega}{f}r^{2z-2}\partial_r\left(r^{2-\theta}h_{ty}-k\int_{r_c}^r ds\  s^{3-z-\theta}a_y\right)=0\ .
\end{equation}
This is structurally similar to \eqref{dg-upstairseom3} in terms of 
a new field variable
\begin{equation}\label{tilde-hty}
r^{2-\theta}\tilde{h}_{ty}=r^{2-\theta}h_{ty}-k\int_{r_c}^r ds\  s^{3-z-\theta}a_y\ .
\end{equation}
At the boundary $r=r_c\sim 0$, we impose the boundary conditions that 
these perturbations vanish, as done previously.
This in turn motivates a redefinition to new field variables in the
$y$-compactified theory as 
\begin{equation}\label{fld-redef}
\begin{aligned}
\tilde{\scriptA}_t&=\scriptA_t-k\int_{r_c}^{r}ds\,s^{3-z-\theta}\chi\ ,\\
\tilde{\scriptA}_x&=\scriptA_x\ .
\end{aligned}
\end{equation}
For the new gauge field variables $\scriptA_t$ and $\scriptA_x$, we
define the field strengths $\tilde{\scriptF}_{rt}$ and
$\tilde{\scriptF}_{tx}$ as (in radial gauge
$\tilde{\scriptA}_r=\scriptA_r=0$)
\begin{equation}
\tilde{\scriptF}_{rt}=\scriptF_{rt}-kr^{3-z-\theta}\chi\ , \qquad
\tilde{\scriptF}_{tx}=\partial_t\scriptA_x-\partial_x\tilde{\scriptA}_t\ ,\qquad
\tilde{\scriptF}_{rx}=\scriptF_{rx}=\partial_r\scriptA_x\ .
\end{equation}
In terms of the newly defined field strengths, the Maxwell's Equations
\eqref{nu-t-eqn}-\eqref{nu-r-eqn}, Bianchi identity \eqref{bianchi-1}
and the equation of motion for $\chi$ \eqref{down-scalareqn} become
\begin{eqnarray}\label{eom-redhty}
\partial_r(r^{z+\theta-3}\tilde{\scriptF}_{rt})-\frac{r^{z+\theta-3}}{f}\partial_x\left(\tilde{\scriptF}_{tx}-k\int_{r_c}^{r}ds\,s^{3-z-\theta}\partial_x\chi\right)&=&0\ ,\\
\label{eom-redhxy}
\partial_r(r^{-1-z+\theta}f\scriptF_{rx})-\frac{r^{z+\theta-3}}{f}\partial_t\left(\tilde{\scriptF}_{tx}-k\int_{r_c}^{r}ds\,s^{3-z-\theta}\partial_x\chi\right)&=&0\ ,\\
\label{eom-redhry}
\partial_t\tilde{\scriptF}_{rt}-r^{2-2z}f\partial_x\scriptF_{rx}&=&0\ ,\\
\label{red-bianchi}
\partial_t\scriptF_{rx}+\partial_x\tilde{\scriptF}_{tr}-\partial_r\tilde{\scriptF}_{tx}&=&0\ ,\\
\label{eom-redchi}
\partial_r(r^{5-z-\theta}f\partial_r\chi)-k^2r^{3-z-\theta}\chi-\frac{r^{3+z-\theta}}{f}\partial_t^2\chi+r^{5-z-\theta}\partial_x^2\chi-k\tilde{\scriptF}_{rt}&=&0\ .
\end{eqnarray}
Differentiating \eqref{eom-redhry} w.r.t. $t$ we can eliminate
$\scriptF_{rx}$ using the Bianchi Identity \eqref{red-bianchi} to get
the following equation
\begin{equation}\label{ME1}
\partial_t^2\tilde{\scriptF}_{tr}+r^{2-2z}f\partial_x(-\partial_x\tilde{\scriptF}_{tr}+\partial_r\tilde{\scriptF}_{tx})=0\ .
\end{equation}
In the the near horizon region approximating the thermal factor as
$f(r) \approx (2+z-\theta)\frac{(1/r_0)-r}{1/r_0}$ and
parametrizing the frequency as $\omega = -i\Gamma$ for some positive
$\Gamma$ so that the perturbations decay in time, \eqref{ME1}
can be written as
\begin{equation}
\label{t-der-diff}
\left(1+(2+z-\theta) r_0^{2z-2}\frac{q^2}{\Gamma^2}\cdot \frac{\frac{1}{r_0}-r}{\frac{1}{r_0}}\right)\tilde{\scriptF}_{tr}\approx-(2+z-\theta)r_0^{2z-2}\frac{iq}{\Gamma^2}\cdot\frac{\frac{1}{r_0}-r}{\frac{1}{r_0}}\partial_r\tilde{\scriptF}_{tx}\ .
\end{equation} 
Assuming the bound \eqref{outerbound}, we differentiate both sides 
w.r.t. $x$ and approximate \eqref{t-der-diff} further
\begin{equation}\label{ftrestimate}
\partial_x\tilde{\scriptF}_{tr}\approx (2+z-\theta)\frac{q^2r_0^{2z-2}}{\Gamma^2}\cdot \frac{\frac{1}{r_0}-r}{\frac{1}{r_0}}\partial_r\tilde{\scriptF}_{tx}
\ \equiv\ \epsilon (2+z-\theta) \partial_r\tilde{\scriptF}_{tx}\ ,
\end{equation}
where 
\begin{equation}\label{assumption1}
\epsilon =\frac{q^2}{\Gamma^2}r_0^{2z-2}\cdot\frac{\frac{1}{r_0}-r}{\frac{1}{r_0}}\ll 1\ ,
\end{equation}
which is essentially implied by \eqref{outerbound}.
In other words, we have 
\begin{equation}\label{fieldstrcomparison}
\partial_x \tilde{\scriptF}_{tr}\ll \partial_r \tilde{\scriptF}_{tx}\ ,
\end{equation}
 which in turn simplifies the Bianchi Identity to
\begin{equation}\label{approx-bianchi}
\partial_t \scriptF_{rx} = \partial_x\tilde{\scriptF}_{rt}+\partial_r \tilde{\scriptF}_{tx}\ \sim\ \partial_r \tilde{\scriptF}_{tx}\ .
\end{equation}

Differentiating \eqref{eom-redhxy} w.r.t $t$ 
we get
\begin{equation}
\partial_r(r^{\theta-z-1}f \partial_t\scriptF_{rx})-\frac{r^{z+\theta-3}}{f}\partial_t^2 \tilde{\scriptF}_{tx}+k\frac{r^{z+\theta-3}}{f}\int_{r_c}^r s^{3-z-\theta} \cdot \partial_t^2\partial_x \chi(s)\ ds=0\ .
\end{equation}
Using the approximate Bianchi identity \eqref{approx-bianchi}, to
substitute for $\scriptF_{rx}$ and then multiplying throughout with
$-\frac{f}{r^{z+\theta-3}}$ we get a sourced wave equation for the
field strength $\scriptF_{tx}$
\begin{equation}
\label{waveequation-ftx}
\partial_t^2 \tilde{\scriptF}_{tx}-\nu^2\left(\frac{1}{r_0}-r\right)\partial_r \left(\left(\frac{1}{r_0}-r\right)\partial_r \tilde{\scriptF}_{tx} \right)\approx k \int_{r_c}^{r}s^{3-z-\theta}\cdot \partial_t^2 \partial_x \chi(s)ds\ ,
\end{equation}
with $\nu$ in \eqref{nu}.\\
Likewise for the scalar equation of motion \eqref{eom-redchi}, using
the approximation \eqref{outerbound} we can drop the term involving 
$\partial_x^2\chi$ compared to the other terms: thus in the near
horizon regime we obtain
\begin{equation}\label{waveequation-chi}
\partial_t^2 \chi -\nu^2\left(\frac{1}{r_0}-r\right)\partial_r \left(\left(\frac{1}{r_0}-r\right)\partial_r \chi \right)+\frac{\nu^2k^2 r_0}{2+z-\theta}\left(\frac{1}{r_0}-r\right)\chi = -\nu k r_0^{4-\theta}\left(\frac{1}{r_0}-r\right)\tilde{\scriptF_{rt}}\ .
\end{equation}
that the first term in \eqref{waveequation-chi} is sub-dominant than 
the third term by a factor of $\frac{\Gamma^2}{\frac{1}{r_0}-r} \ll 1$. 
Thus the leading order behaviour for the scalar field $\chi$ can 
simply be estimated as
\begin{equation}
\label{chi-estimate}
\chi^{(0)}\ \approx\ -\frac{2+z-\theta}{\nu k}r_0^{3-\theta}
\tilde{\scriptF}_{rt}^{(0)}\ ,
\end{equation}
where the superscript $(0)$ is the leading order behaviour of the
field $\chi$ at $q=\Gamma=0$ since we have explicitly dropped the
subleading derivative terms. Now, in the near horizon regime, we can 
use (\ref{ftrestimate}) to find\
$\del_x\chi^{(0)}\sim \del_x \tilde{\scriptF}_{tr}^{(0)} \sim 
\epsilon \del_r\tilde{\scriptF}_{tx}^{(0)}$.\ Using this, we can 
estimate the right hand side of \eqref{waveequation-ftx} as 
\begin{equation}\label{chi-estimate2}
k \int_{r_c}^{r}s^{3-z-\theta}\cdot \partial_t^2 \partial_x \chi(s)ds \approx k \int_{r_c}^r r_0^{z+\theta-3}\partial_t^2 \partial_x \chi
\sim \partial_t^2 \int_{r_c}^r \epsilon\ \partial_s \tilde{\scriptF}_{tx} ds
\sim \epsilon \cdot \partial_t^2 \tilde{\scriptF}_{tx}\ .
\end{equation}
What this means is that while the gauge field perturbation $a_y$ (or 
$\chi$) is not subleading to $h_{ty}$ (or $\scriptA_t$), once we 
incorporate its effects in terms of the variable ${\tilde h_{ty}}$ 
(or $\tilde{\scriptA}_t$) the remaining contributions are in fact 
subleading, as we see here in (\ref{chi-estimate2}).\\
The above estimate implies that upto leading order, 
\eqref{waveequation-ftx} is in fact a source free wave equation whose 
ingoing solution is
\begin{equation}
\tilde{\scriptF}_{tx} = f_1\left(t+\frac{1}{\nu}\log\left(\frac{1}{r_0}-r\right)\right)\ ,
\end{equation}
which further implies
\begin{equation}
\partial_t\tilde{\scriptF}_{tx}+\nu\left(\frac{1}{r_0}-r\right)\partial_r\tilde{\scriptF}_{tx}=0\ .
\end{equation}
Using \eqref{approx-bianchi}, we can write the above expression as a 
perfect derivative in $t$, \ie\ 
\be
\del_t \left(\tilde{\scriptF}_{tx}+\nu\Big(\frac{1}{r_0}-r\Big)
\tilde{\scriptF}_{rx}\right) = 0\ .
\ee
 Imposing the boundary condition that
the solutions decay as $t \rightarrow \infty$ we end up with the
following relation
\begin{equation}\label{ftx-frx-eqn}
\tilde{\scriptF}_{tx}+\nu\left(\frac{1}{r_0}-r\right)\scriptF_{rx}=0\ .
\end{equation}
We can derive this result alternatively arguing as follows, looking for 
an ingoing condition as in \eqref{infallH}. In this case we have 
identified $\tilde{\scriptA}_t$ as the relevant perturbative mode. We 
can write the newly defined field strengths in terms of $h_{ty}$, $h_{xy}$ 
and $a_y$ along the lines of \eqref{fldstrngths} as
\begin{equation}
\partial_t \tilde{\scriptF}_{tx}=- r^{2-\theta}\omega(\omega h_{xy}+q \tilde{h}_{ty})\ ,\qquad 
\scriptF_{rx}=\partial_r (r^{2-\theta}h_{xy})=\frac{r^{z+1-\theta}}{f}H\ .
\end{equation}
From \eqref{eom-redhxy}, we have
\begin{equation}\label{delrH}
  \partial_r H= \frac{r^{z+\theta-3}}{f}\partial_t(\tilde{\scriptF}_{tx}-k\int_{r_c}^{r}ds\,s^{3-z-\theta}\partial_x a_y)\ .
\end{equation}
From \eqref{chi-estimate2} cancelling the $\partial_t^2 \equiv \Gamma^2$ 
factor throughout, it follows that
\begin{equation}\label{chi-estimate3}
  k\int_{r_c}^r ds\,s^{3-z-\theta}\partial_x a_y\sim \epsilon \tilde{\scriptF}_{tx}\ .
\end{equation}
Substituting this equation in \eqref{delrH}, we get
\begin{equation}
  \partial_r H\approx \frac{r^{z+\theta-3}}{f}\partial_t\tilde{\scriptF}_{tx}\ .
\end{equation}
We expect on physical grounds that the ingoing condition on $H$ defined 
in terms of $h_{xy}$ is still the same as \eqref{infallH} in the case 
without the gauge field since this shear mode is expected to be 
ingoing: this gives
\begin{equation}
  \partial_r H=-r^{\theta-2}\partial_t \scriptF_{rx}\ .
\end{equation}
(In the above equations, we have used the $y$-compactified variables 
and higher dimensional ones in the same equations, with the understanding 
that they are interchangeable from the context.)
Equating the two expressions for $\partial_r H$ above we recover
\eqref{ftx-frx-eqn}, which is analogous to (\ref{dg-ftx-frx-reln}) 
in the case without the gauge field. This vindicates our intuition 
on using the $\tilde{\scriptA}_t,\ \tilde{\scriptA}_x$ field variables 
to obtain the diffusion equation here with the gauge field.

Along the lines of \eqref{horizoncurrent1}, \eqref{horizoncurrent2}, 
we define the currents in the new tilde variables as
\begin{equation}\label{def-current}
j^x=n_r\scriptF^{xr}=\frac{\scriptF_{rx}}{g_{xx}\sqrt{g_{rr}}}\ , \qquad \tilde{j}^t=n_r\tilde{\scriptF}^{tr}=\frac{\tilde{\scriptF}_{rt}}{g_{tt}\sqrt{g_{rr}}}\ ,
\end{equation}
since as we have seen, these ${\tilde\scriptA_\mu}$ variables play the 
role here of the earlier variables $\scriptA_\mu$\ (it would be 
interesting to find appropriate modifications of the prescriptions in 
\cite{Iqbal:2008by} here).
At this point we make another assumption, namely
\begin{equation}\label{assumption2}
|\partial_t \scriptA_x| \ll |\partial_x \tilde{\scriptA}_t|\ ,
\end{equation}
which is very similar to (\ref{ansatz2}) but for the 
${\tilde \scriptA}_t, {\tilde \scriptA}_x$ variables (\ref{fld-redef}). 
This implies
\begin{equation}\label{ftx-approx}
\tilde{\scriptF}_{tx} \approx -\partial_x \tilde{\scriptA}_t\ .
\end{equation}
We can now formulate Fick's Law \ie\ $j^{x}=-\mathcal{D}\partial_x
\tilde{j}^{t}$ on the stretched horizon and calculate the diffusion
constant as
\begin{align}
\mathcal{D}\equiv -\frac{j^x}{\partial_x\tilde{j}^t}&=-\frac{g_{tt}}{g_{xx}}\frac{\scriptF_{rx}}{\partial_x\tilde{\scriptF}_{rt}} \approx -r_0^{z-1}\frac{\tilde{\scriptF}_{tx}}{\partial_x\tilde{\scriptF}_{rt}} \ ,
\end{align}
where we have used \eqref{ftx-frx-eqn} to write the third equality.
Using \eqref{ftx-approx}, the diffusion constant at leading order 
is given by
\begin{equation}\label{diffconstant}
\mathcal{D}=r_0^{z-1}\left.\frac{\tilde{\scriptA}_t}{\tilde{\scriptF}_{rt}}\right|_{r \sim r_h}\ ,
\end{equation} 
where $r_h$ is the location of the stretched horizon, and the prefactor
arises from the metric factors as in (\ref{diffconst}).

\subsubsection{Shear diffusion constant: $z<4-\theta$}

Making an ansatz of the form \eqref{up-seriesansatz} naturally implies
such a series expansion ansatz for the fields ${\tilde \scriptA}_t$,
${\tilde \scriptA}_x$ and $\chi$ in the $y$-compactified theory.
\begin{equation}\label{down-seriesansatz}
\begin{pmatrix}
\tilde{\scriptA}_{t}(t,x,r)\\
\tilde{\scriptA}_{x}(t,x,r)\\
\chi(t,x,r)
\end{pmatrix}
\equiv
\begin{pmatrix}
\tilde{\scriptA}_{t}^{(0)}(t,x,r)\\
\tilde{\scriptA}_{x}^{(0)}(t,x,r)\\
\chi^{(0)}(t,x,r)
\end{pmatrix}+ 
\begin{pmatrix}
\tilde{\scriptA}_{t}^{(1)}(t,x,r)\\
\tilde{\scriptA}_{x}^{(1)}(t,x,r)\\
\chi^{(1)}(t,x,r)
\end{pmatrix}+ \cdots\ , \qquad \begin{pmatrix}
\tilde{\scriptA}_{t}^{(1)}(t,x,r)\\
\tilde{\scriptA}_{x}^{(1)}(t,x,r)\\
\chi^{(1)}(t,x,r)
\end{pmatrix}= O\left(\frac{q^2}{T^{2/z}}\right)\ .
\end{equation}
The $q=\Gamma=0$ sector of \eqref{eom-redhty} which is
\begin{equation}
\partial_r(r^{z+\theta-3}\partial_r\tilde{\scriptA}_{t})=0\ ,
\end{equation}
gives us an expression for the leading solution of $\tilde{\scriptA}_t$ 
\begin{equation}\label{tilde-At-integral}
\tilde{\scriptA}_{t}^{(0)}(t,x,r)=C\ e^{-\Gamma t+ iq x}\int_{r_c}^r dr.\ r^{3-z-\theta}\ ,
\end{equation}
where $C$ is an arbitrary constant. When $4-z-\theta>0$ the leading solution 
$\tilde{\scriptA}_{t}^{(0)}$ has a power-law behaviour 
\begin{equation}\label{leading-tildeAt}
\tilde{\scriptA}_t^{(0)}(t,x,r)=e^{-\Gamma t+ iq x}\frac{C}{(4-z-\theta)}r^{4-z-\theta}
\end{equation}
It is expected that close to the boundary \ie\ near $r \approx r_c$
the hyperscaling violating phase breaks down and we require $r_0 r_c
\ll 1$. The analogous statement for the boundary field theory will be
to assume that the temperature is sufficiently below the UV
cut-off. Thus, the condition $z< 4-\theta$ arises from the boundary
condition that $\tilde{\scriptA}_t^{(0)} \rightarrow 0$ as $r
\rightarrow 0$.

Substituting $\tilde{\scriptA}_t^{(0)}$ in \eqref{down-scalareqn}, the
particular solution to the inhomogeneous equation (at $q=0$,
$\omega=0$) for $\chi$ is
\begin{equation}\label{leadingchi}
\chi^{(0)}=-\frac{C}{k}\ .
\end{equation}
Substituting $\chi^{(0)}=a_y^{(0)}=-C/k$ in \eqref{tilde-hty}, and
considering only the leading order terms we get
\begin{equation}\label{tildehtysoln}
\tilde{h}_{ty}^{(0)}=h_{ty}^{(0)}+\frac{C}{(4-z-\theta)}r^{2-z}\ .
\end{equation}
Thus, we see that although $h_{ty}=r^{2-z}$ does not satisfy the
linearized equations \eqref{upstairseom1}-\eqref{upstairseom4} at
$q=0$, $\omega=0$, the $r^{2-z}$ fall-off appears in the expression
for $\tilde{h}_{ty}$ which is indeed the relevant perturbative mode
that should be considered. We see that
$\tilde{h}_{ty}=\frac{C}{4-z-\theta}r^{2-z}$ and $a_y=-\frac{C}{k}$
indeed satisfy the linearized equations
\eqref{eom-redhty}-\eqref{eom-redchi} at $q=0$, $\omega=0$. Note that 
this implies that the solutions of interest here in the original 
variables are $h_{ty}^{(0)}=0$ and $a_y^{(0)}=-{C\over k}$, as can be
seen from the form of $\tilde{h}_{ty}$. Thus the solutions of 
relevance arise entirely from the leading solution to the gauge 
field perturbation. It is important to note that the solution 
$a_y^{(0)}=const$ does not change the asymptotic boundary conditions 
on the background being hyperscaling violating Lifshitz.

The leading solution for $\scriptA_x$ \ie\ $\scriptA_x^{(0)}$ can be
determined by plugging in the series ansatz for $\scriptA_{x}$ and
$\tilde{\scriptA}_t$ in \eqref{eom-redhxy}. The leading order equation
is given by
\begin{equation}
\partial_r\scriptA_x^{(0)}=\frac{i\Gamma}{q}\frac{r^{2z-2}}{f}\partial_r\tilde{\scriptA}_t^{(0)}\ .
\end{equation}
Integrating the above and using \eqref{leading-tildeAt} we obtain 
an expression for $\scriptA_x^{(0)}$ as
\begin{equation}\label{Ax-soln}
\scriptA_x^{(0)}=\frac{i\Gamma}{q}\frac{Ce^{-\Gamma t+iqx}}{(2+z-\theta)r_0^{2+z-\theta}}\log(1-(r_0r)^{2+z-\theta})\ .
\end{equation}
From \eqref{tilde-At-integral} and the solution derived above, we see
that the assumption \eqref{assumption2} is essentially
\begin{equation}\label{innerbound1}
\frac{\Gamma^2}{q^2}r_0^{2-2z}\log\Big(\frac{(1/r_0)}{(1/r_0)-r_h}\Big)\ll 1\ .
\end{equation}
Using $\frac{\Gamma}{q}\sim \frac{q}{r_0^{2-z}}$ and noting that the
temperature $T \sim r_0^z$, we can recast this condition as
\begin{equation}\label{innerbound2}
\frac{q^2}{T^{2/z}}\log\Big(\frac{(1/r_0)}{(1/r_0)-r_h}\Big)\ll 1\ .
\end{equation}
Physically the above assumption means that we cannot push the
stretched horizon located at $r_h$ exponentially close to the horizon
$\frac{1}{r_0}$ as before, in (\ref{combinedcriteria}).

Using \eqref{leading-tildeAt} we can now evaluate the shear diffusion
constant on the stretched horizon for the hyperscaling violating
theory with $4-z-\theta >0$ as
\begin{equation}
\mathcal{D}=r_0^{z-1}\cdot \frac{1}{(4-z-\theta)r_h} \approx \frac{r_0^{z-2}}{4-z-\theta}+O(q^2)\ .
\end{equation}
The solution for $\tilde{\scriptA}_t^{(0)}$ is evaluated at the
stretched horizon $r_h$: however $r_h \sim \frac{1}{r_0}+O(q^2)$ so to
leading order $\mathcal{D}$ is evaluated at the horizon
$\frac{1}{r_0}$. It is interesting that the effect of the hyperscaling
violating exponent $\theta$ cancels in the final expression for
$\mathcal{D}$ which is essentially the ratio of $\tilde{\scriptA}_t$
to a field strength $\tilde{\scriptF}_{rt}$ both of which has
non-trivial $\theta$-dependence.

Using the expression \eqref{hvtemperature} we can express the
diffusion constant in terms of the temperature as
\begin{equation}\label{DT(2-z)/z:ay}
\mathcal{D}= \frac{1}{4-z-\theta}\left(\frac{4\pi}{2+z-\theta}\right)^{\frac{z-2}{z}}T^{\frac{z-2}{z}}
\end{equation}
which is identical to the one obtained in \cite{Kolekar:2016pnr} for
the case without the gauge field, for $d_i=2$ spatial dimensions. As 
discussed there, for pure $AdS$ when $z=1, \theta=0$, we recover the 
standard relation $\mathcal{D}=\frac{1}{4\pi T}$ which further implies
$\frac{\eta}{s}=\frac{1}{4\pi}$. Likewise for all theories with 
$z=1$, it can be seen that $\theta$ cancels from the prefactors in 
$\mathcal{D}$ which becomes $\mathcal{D}=\frac{1}{4\pi T}$~. This is 
in accord with the known behaviour \cite{Kovtun:2003wp} of \eg\ 
nonconformal $Dp$-branes whose dimensional reduction on the 
transverse sphere $S^{8-p}$ gives rise to hyperscaling violating 
theories with $z=1,\ \theta\neq 0$ \cite{Dong:2012se}: it would seem
reasonable to expect that the sphere should not affect long-wavelength 
diffusive properties.

\subsubsection{Shear diffusion constant: $z=4-\theta$}

Now, we focus on the family of hyperscaling violating solution where
$z=4-\theta$. In this case, from \eqref{tilde-At-integral} it follows
that the leading solution of ${\tilde \scriptA}_t$ has logarithmic behaviour
\begin{equation}\label{tilde-At-special}
{\tilde \scriptA}_t^{(0)}=Ce^{-\Gamma t +i qx}\log \frac{r}{r_c}\ , 
\qquad \quad z=4-\theta\ .
\end{equation}
Working further, we can evaluate the diffusion constant upto leading
order from \eqref{diffconstant} as
\begin{equation}
\mathcal{D} = r_0^{z-2}\log \frac{1}{r_0r_c}\ .
\end{equation}
This implies that in the low temperature limit as $r_0 \rightarrow 0$,
the diffusion constant vanishes if $z >2$. The new condition on the
exponents $z$ and $\theta$, namely $z< 4-\theta$ appears to be a new
constraint which is separate from the null energy conditions
\begin{equation}
(2-\theta)(2(z-1)-\theta)\geq 0\ , \qquad \quad (z-1)(2+z-\theta)\geq 0\ .
\end{equation}
The regime of validity for this analysis (equivalently, the
``thickness'' of the stretched horizon) gets modified in this special
case to
\begin{equation}\label{assumption-specialcase}
\exp\left(-\frac{T^{2/z}}{q^2}\frac{1}{\log\frac{1}{r_0r_c} }\right)\ll\frac{\frac{1}{r_0}-r_h}{\frac{1}{r_0}}\ll \frac{q^2}{T^{2/z}}\log^2 \frac{1}{r_0r_c}\ .
\end{equation}
However, since we are manifestly in the hydrodynamic regime, it means
$r_c \ll \frac{1}{r_0}$ implying $\log \frac{1}{r_0r_c} \gg 1$. This
does not over-constrain the window of the stretched horizon: however
the subleading terms contain the logarithmic piece affecting the
validity of the series expansion.

The logarithmic scaling necessitates the presence of the UV scale
$r_c$ appearing in the diffusion constant in the hydrodynamic
description which is manifestly a description at long
wavelengths. However from our discussion, it is clear that this is due
to the two fall-offs for $\tilde{\cal A}_t$ coinciding when
$z=4-\theta$: this leads to the second solution being logarithmic and
thence to the scaling above in $\mathcal{D}$.  Recall that the
parameters $z$ and $\theta$ are related precisely in this way when the
hyperscaling violating theory is constructed from the
$x^{+}$-reduction of $AdS$ plane waves (or highly boosted $AdS_5$
black branes), as well as nonconformal $Dp$-brane plane waves, as
discussed in \cite{Kolekar:2016pnr}. (The zero temperature $AdS$ plane 
waves are structurally similar to the null deformations appearing 
in the string realizations \cite{Balasubramanian:2010uk,Donos:2010tu} 
of $z=2$ Lifshitz theories, except that the null deformation is
normalizable.) As outlined in \cite{Kolekar:2016pnr}, to gain more
insight into the diffusion behaviour, it might be interesting to
understand the null reduction of the boosted black brane and its
hydrodynamics in greater detail. This might be similar in spirit to
nonconformal brane hydrodynamics arising under dimensional reduction
of the hydrodynamics of black branes in M-theory
\cite{Kanitscheider:2008kd,Kanitscheider:2009as}, although the details
are likely to be interestingly different of course. It is also worth
noting that in the higher dimensional description, these D-brane plane
waves are dual to excited states in the field theory which correspond
to anisotropic phases in the boosted frame: the corresponding
anisotropic hydrodynamics might be interesting as well\ (see \eg\
\cite{Mateos:2011tv,Erdmenger:2011tj,Rebhan:2011vd,Polchinski:2012nh,
  Mamo:2012sy,Giataganas:2013hwa,
  Jain:2014vka,Jain:2015txa} for previous studies of
anisotropic systems and shear viscosity, and \eg\
\cite{Cremonini:2011iq} for a review of the viscosity bound and
violations).

\subsection{Subleading terms for $z<4-\theta$}

In this section we will estimate the subleading terms as proposed in
\eqref{down-seriesansatz} and explicitly show that
$\tilde{\scriptA}_t^{(1)}$, $\tilde{\scriptA}_x^{(1)}$ and
$\chi^{(1)}$ (infact all the other terms following it) are subleading
compared to the leading order values $\tilde{\scriptA}_t^{(0)}$,
$\tilde{\scriptA}_x^{(0)}$ and $\chi^{(0)}$ respectively.

\subsubsection*{Estimate for $\tilde{\scriptA}_t^{(1)}$}

Substituting the series for $\tilde{\scriptA}_t^{(1)}$ from
\eqref{down-seriesansatz} in \eqref{eom-redhty}, we get
\begin{equation}
\partial_r(r^{z+\theta-3}(\partial_r\tilde{\scriptA}_t^{(0)}+\partial_r\tilde{\scriptA}_t^{(1)})+\cdots)-\frac{r^{z+\theta-3}}{f}\partial_x\left(\tilde{\scriptF}_{tx}^{(0)}-k\int_{r_c}^{r}ds\,s^{3-z-\theta}\partial_x\chi^{(0)}+\cdots\right)=0\ .
\end{equation}
The leading term in the above equation is
$\partial_r(r^{z+\theta-3}\partial_r\tilde{\scriptA}_t^{(0)})=0$,
which is consistent with \eqref{leading-tildeAt}. $O(q^2)$ terms in
the above equation give
\begin{equation}
\partial_r(r^{z+\theta-3}\partial_r\tilde{\scriptA}_t^{(1)})-\frac{r^{z+\theta-3}}{f}\partial_x\left(\partial_t\scriptA_x^{(0)}-\partial_x\tilde{\scriptA}_t^{(0)}\right)=0\ .
\end{equation}
Here we have neglected $k\int_{r_c}^{r}ds\,s^{3-z-\theta}\partial_x\chi^{(0)}$ since
$k\int_{r_c}^{r}ds\,s^{3-z-\theta}\partial_x\chi^{(0)}\ll
\tilde{\scriptF}_{tx}^{(0)}$, using the arguments in \eg\ 
(\ref{chi-estimate}), (\ref{chi-estimate2}). Then
\begin{equation}
\partial_r \tilde{\scriptA}_t^{(1)} \sim \frac{1}{r_0} \left(q^2 \log \left( \frac{1/r_0}{1/r_0 -r}\right) + \frac{\Gamma^2}{r_0^{2(z-1)}}\log^2 \left( \frac{1/r_0}{1/r_0 -r} \right)^2\right)\tilde{\scriptA}_t^{(0)}\ .
\end{equation}
Using the estimate $\frac{\Gamma}{q}\sim \frac{q}{T^{2/z-1}}$, we can write
\begin{equation}\label{At1-r-der}
\partial_r \tilde{\scriptA}_t^{(1)} \sim r_0 \left[\frac{q^2}{T^{2/z}}\log \left(\frac{\frac{1}{r_0}}{\frac{1}{r_0}-r}\right) + \frac{q^4}{T^{4/z}}\log^2 \left(\frac{\frac{1}{r_0}}{\frac{1}{r_0}-r}\right) \right]\tilde{\scriptA}_t^{(0)}\ .
\end{equation}
Integrating the above equation,
\begin{equation}\label{tildeAt1}
\tilde{\scriptA}_t^{(1)} \sim -(1-r_0r)\Big[\frac{q^2}{T^{2/z}}\Big(1+\log\Big(\frac{\frac{1}{r_0}}{\frac{1}{r_0}-r}\Big)\Big) +\frac{q^4}{T^{4/z}}\Big(1+\log\Big(\frac{\frac{1}{r_0}}{\frac{1}{r_0}-r}\Big)+\log^2\Big(\frac{\frac{1}{r_0}}{\frac{1}{r_0}-r}\Big)\Big)\Big]\tilde{\scriptA}_t^{(0)}\ ,
\end{equation}
which implies $\tilde{\scriptA}_t^{(1)}\ll \tilde{\scriptA}_t^{(0)}$.

\subsubsection*{Estimate for $\scriptA_x^{(1)}$}
Substituting the series ansatz for $\scriptA_x$ i.e \eqref{down-seriesansatz} in \eqref{eom-redhxy} gives
\begin{equation}
\partial_r\scriptA_x^{(0)}+\partial_r\scriptA_x^{(1)}+\cdots=\frac{i\Gamma}{q}\frac{r^{2z-2}}{f}(\partial_r\tilde{\scriptA}_t^{(0)}+\partial_r\tilde{\scriptA}_t^{(1)}+\cdots)\ .
\end{equation}
The leading terms have been derived in \eqref{Ax-soln}, so we will focus on $O(q^2)$ terms which gives us the equation
\begin{equation}
\partial_r\scriptA_x^{(1)}=\frac{i\Gamma}{q}\frac{r^{2z-2}}{f}\partial_r\tilde{\scriptA}_t^{(1)}\ ,
\end{equation}
which give
\begin{equation}
\scriptA_x^{(1)} \sim \Big(\frac{q^2}{T^{2/z}}\log^2\Big(\frac{1/r_0}{1/r_0 -r}\Big)+\frac{q^4}{T^{4/z}}\log^3 \Big(\frac{1/r_0}{1/r_0 -r} \Big)^2\Big)\frac{i\Gamma r_0^{2-2z}}{q}\tilde{\scriptA}_t^{(0)}\ .
\end{equation}
Using $\scriptA_x^{(0)}\sim \frac{i\Gamma r_0^{2-2z}}{q}\log\left(\frac{1/r_0}{1/r_0 -r}\right)\tilde{\scriptA}_t^{(0)}$,
\begin{equation}
\scriptA_x^{(1)} \sim \Big[\frac{q^2}{T^{2/z}}\log\Big( \frac{1/r_0}{1/r_0 -r}\Big)+\frac{q^4}{T^{4/z}}\log^2\Big( \frac{1/r_0}{1/r_0 -r}\Big) \Big]\scriptA_x^{(0)}\ ,
\end{equation}
which implies $\scriptA_x^{(1)}\ll \scriptA_x^{(0)}$.

\subsubsection*{Estimate for $\chi^{(1)}$}
Finally, substituting the series ansatz for $\chi$ i.e \eqref{down-seriesansatz} in \eqref{eom-redchi}, we get
\begin{equation}\begin{split}
\partial_r(r^{5-z-\theta}f(\partial_r \chi^{(0)}+\partial_r \chi^{(1)}))-k^2r^{3-z-\theta}(\chi^{(0)}+\chi^{(1)})&-\frac{r^{3+z-\theta}}{f}(\partial_t^2\chi^{(0)}+\partial_t^2\chi^{(1)})+\cdots \\
&= k \partial_r(\tilde{\scriptA}_t^{(0)}+\tilde{\scriptA}_t^{(1)}+\cdots)\ .
\end{split}\end{equation}
Writing down \eqref{eom-redchi} collecting all $O(q^2)$ terms give
\begin{equation}\label{chi1eqn}
\partial_r(r^{5-z-\theta}f \partial_r \chi^{(1)})-k^2r^{3-z-\theta}\chi^{(1)}=\frac{r^{3+z-\theta}}{f}\Gamma^2 \chi^{(0)}+k\partial_r\tilde{\scriptA}_{t}^{(1)}\ .
\end{equation}
To see that $\chi^{(1)}$ is subleading compared to $\chi^{(0)}$ quickly, let us focus on the first term on both sides of the above equation near the horizon;
\begin{equation}
\partial_r\Big(r_0\Big(\frac{1}{r_0}-r\Big)\partial_r \chi^{(1)}\Big)\sim \frac{q^4}{r_0^2}\frac{1}{r_0(\frac{1}{r_0}-r)}\chi^{(0)}\ ,
\end{equation}
where we have used $\frac{\Gamma}{q}\sim \frac{q}{r_0^{2-z}}$. Integrating twice, we get
\begin{equation}
\chi^{(1)}\sim \frac{q^4}{r_0^4}\log^2\Big(\frac{1/r_0}{1/r_0-r}\Big)\chi^{(0)}\ .
\end{equation}
Using \eqref{innerbound2}, the above expression shows that
$\chi^{(1)}\ll \chi^{(0)}$. This succinct order of magnitude analysis for
the subleading nature of $\chi^{(1)}$ can be substantiated through a 
more detailed analysis as follows. In the near horizon region, 
\eqref{chi1eqn} simplifies to
\begin{equation}
\partial_r\Big(\Big(\frac{1}{r_0}-r\Big)\partial_r \chi^{(1)}\Big)-2(z-1)r_0 \chi^{(1)}=\frac{r_0^{4-z-\theta}}{2+z-\theta}\Big(k\partial_r \tilde{\scriptA}_t^{(1)}+\frac{r_0^{\theta-z-3}}{(2+z-\theta)r_0\Big(\frac{1}{r_0}-r\Big)}\Gamma^2 \chi^{(0)}\Big)\;.
\end{equation}
The Green's function for the above equation is effectively the function $G(r,s)$ that satisfies the equation
\begin{equation}
\partial_r\left(\left(\frac{1}{r_0}-r\right)\partial_r G(r,s)\right)-2(z-1)r_0 \cdot G(r,s)=\delta(r-s)\ .
\end{equation}
The inhomogeneous solution to the Green's function is given by
\begin{equation}
\begin{split}
G_{in}(r,s)= 2 \Theta(r-s)&\big[I_0(2\sqrt{2(z-1)}\sqrt{1-r_0s})\cdot K_0(2\sqrt{2(z-1)}\sqrt{1-r_0r})\\
&-K_0(2\sqrt{2(z-1)}\sqrt{1-r_0s})\cdot I_0(2\sqrt{2(z-1)}\sqrt{1-r_0r})\big]\ .
\end{split}
\end{equation}
Correspondingly the inhomogeneous solution to $\chi^{(1)}$ is given by
\begin{equation}
\chi^{(1)}=\int_0^{1/r_0}ds \cdot G_{in}(r,s)\cdot \frac{r_0^{4-z-\theta}}{2+z-\theta}\left(k\partial_r \tilde{\scriptA}_t^{(1)}+\frac{r_0^{\theta-z-3}}{(2+z-\theta)r_0\left(\frac{1}{r_0}-r\right)}\Gamma^2 \chi^{(0)}\right)\ ,
\end{equation}
where $I_{0}$ and $K_{0}$ are modified Bessel functions of the first
and second kind respectively. Since we are interested only in the
near-horizon behaviour for $\chi^{(1)}$. Instead of explicitly
performing the integral exactly and then taking the limit $r
\rightarrow \frac{1}{r_0}$, we will instead approximate the integrand
close to $\frac{1}{r_0}$. Upto leading order the modified Bessel
functions $I_0$ and $K_0$ near $x \approx 0$ are given by
\begin{equation}
I_{0}(x) \approx 1\ , \qquad \qquad K_{(0)} \approx -\log x + \log 2 - \gamma\ ,
\end{equation}
where $\gamma$ is the Euler constant. Close to the horizon, we can
hence approximate the inhomogeneous part of the Green's function as
\begin{equation}
G_{in}(r,s)= \Theta(r-s)\log\left(\frac{1-r_0 s}{1-r_0 r}\right)\ .
\end{equation}
Hence $\chi^{(1)}$ can be simplified using the above approximation
along with \eqref{At1-r-der}
\begin{equation}
\begin{aligned}
\chi^{(1)} &=\int_0^{1/r_0}ds\,G_{in}(r,s)\,\frac{r_0^{4-z-\theta}}{2+z-\theta}\left(k\partial_r \tilde{\scriptA}_t^{(1)}+\frac{r_0^{\theta-z-3}}{(2+z-\theta)r_0\left(\frac{1}{r_0}-s\right)}\Gamma^2 \chi^{(0)}\right)\\
&\sim  \int_0^{1/r_0}ds\,\Theta(r-s)\log\Big(\frac{1-r_0 s}{1-r_0 r}\Big) \Big[r_0 \Big(\frac{q^2}{T^{2/z}}\log \Big(\frac{\frac{1}{r_0}}{\frac{1}{r_0}-s}\Big)
+ \frac{q^4}{T^{4/z}}\log^2 \Big(\frac{\frac{1}{r_0}}{\frac{1}{r_0}-s}\Big) \Big)\tilde{\scriptA}_t^{(0)} \Big]\\
& \hspace*{6cm} + \int_0^{1/r_0}ds\,\Theta(r-s)\log\Big(\frac{1-r_0 s}{1-r_0 r}\Big)\frac{q^4}{T^{4/z}}\frac{1}{\frac{1}{r_0}-s}\chi^{(0)} \ .\\
\end{aligned}
\end{equation}
The above integral can be divided into two parts. One ranging from $0$
to $r$ and another from $r$ to $\frac{1}{r_0}$. The Heaviside Theta
function is non-zero for $r>s$ only. So, the upper bound in the above
integral can simply be replaced with $r$ instead of $1/r_0$.
Simplifying and performing the integral over $s$ we get,
\begin{equation}\label{chi1}
\begin{split}
\chi^{(1)} \sim &\left[\frac{q^2}{T^{2/z}}\Big\{(1-r_0r) -(1-r_0r)\log(1-r_0r)+(1-r_0r)\log^2(1-r_0r)\Big\}\right.\\
&+\frac{q^4}{T^{4/z}}\left\{-(1-r_0r)+(1-r_0r)\log(1-r_0r) -(1-r_0r)\log^2(1-r_0r)\right.\\
&\left.\left.+(1-r_0r)\log^3(1-r_0r)\right\}\right]\tilde{\scriptA}_t^{(0)}+\frac{q^4}{T^{4/z}}\log^2(1-r_0r)\chi^{(0)}\ll \chi^{(0)}\ .\\
\end{split}
\end{equation}
If we now use the two assumptions mentioned earlier i.e \eqref{outerbound}, \eqref{innerbound2} we explicitly see that $\chi^{(1)} \ll \chi^{(0)}$ thus
demonstrating that all subsequent terms in the series are smaller than
the leading piece.

\subsubsection*{Estimate for $h_{ty}^{(1)}$}

Note that in \eqref{up-seriesansatz} we proposed the series expansion
for the modes $h_{ty}$, $h_{xy}$ and $a_y$.  From the definition of
$\tilde{h}_{ty}$ and the series ansatze \eqref{up-seriesansatz},
\eqref{down-seriesansatz}, we can write
\begin{equation}\label{series-eqn}
h_{ty}^{(0)}+h_{ty}^{(1)}+\cdots=\Big(\tilde{h}_{ty}^{(0)}+kr^{\theta-2}\int_{r_c}^{r}ds\,s^{3-z-\theta}a_y^{(0)}\Big)+\Big(\tilde{h}_{ty}^{(1)}+kr^{\theta-2}\int_{r_c}^{r}ds\,s^{3-z-\theta}a_y^{(1)}\Big)+\cdots\ .
\end{equation}
From \eqref{tildeAt1} and \eqref{chi1}, we have
\begin{equation}
h_{ty}^{(1)}=\tilde{h}_{ty}^{(1)}+kr^{\theta-2}\int_{r_c}^{r}ds\,s^{3-z-\theta}a_y^{(1)} \sim O\Big(\frac{q^2}{T^{2/z}}\Big)\tilde{h}_{ty}^{(0)} \ .
\end{equation}
Using
\begin{equation}
\tilde{h}_{ty}^{(2)}\sim \frac{q^{2}}{T^{2/z}} \tilde{h}_{ty}^{(1)}\ , \qquad\qquad a_y^{(2)}\sim \frac{q^{2}}{T^{2/z}} a_y^{(1)}\ ,
\end{equation}
we see that
\begin{equation}
\frac{h_{ty}^{(2)}}{h_{ty}^{(1)}}=\frac{\tilde{h}_{ty}^{(2)}+kr^{\theta-2}\int_{r_c}^{r}ds\,s^{3-z-\theta}a_y^{(2)}}{\tilde{h}_{ty}^{(1)}+kr^{\theta-2}\int_{r_c}^{r}ds\,s^{3-z-\theta}a_y^{(1)}}\sim O\Big(\frac{q^2}{T^{2/z}}\Big)\ll 1 \ .
\end{equation}
Thus we see that the mode $h_{ty}$ also admits a series expansion in
the parameter $\frac{q^2}{T^{2/z}}$ in the near-horizon region. This is 
of course expected from the self-consistent series expansions of 
${\tilde h_{ty}}, h_{xy}, a_y$.

\subsubsection{Subleading terms for $z=4-\theta$}

In this case, from the solutions of $\tilde{\scriptA}_t$ 
\eqref{tilde-At-special} and $\scriptA_x$ \eqref{Ax-soln} we get, 
\begin{equation}
\frac{\scriptA_x^{(0)}}{\scriptA_t^{(0)}}\sim \frac{1}{r_0^{2(z-1)}}\frac{\Gamma}{q}\frac{\log(\frac{1/r_0}{1/r_0-r})}{\log(\frac{1}{r_0r_c})}\ .
\end{equation}
Imposing \eqref{assumption2} then implies 
\begin{equation}
\frac{1}{r_0^{2(z-1)}}\cdot \frac{\Gamma^2}{q^2}\cdot \frac{\log(\frac{1/r_0}{1/r_0-r})}{\log(\frac{1}{r_0r_c})} \ll 1\ .
\end{equation}
We can obtain an estimate for $\mathcal{D}$ in this case from the
diffusion equation which is $\frac{\Gamma}{q} \sim
\frac{q}{T^{2/z-1}}\log(\frac{1}{r_0r_c})$. Thus, the assumptions in
this special case gets modified to \eqref{assumption-specialcase}. The
subleading term for $\tilde{\scriptA}_t$ now is given by
\begin{equation}
\partial_r \tilde{\scriptA}_t^{(1)} \sim r_0 \left[\frac{q^2}{T^{2/z}}\log \left(\frac{\frac{1}{r_0}}{\frac{1}{r_0}-r}\right) + \frac{q^4}{T^{4/z}}\log^2 \left(\frac{\frac{1}{r_0}}{\frac{1}{r_0}-r}\right)\log\left(\frac{1}{r_0r_c}\right) \right]\tilde{\scriptA}_t^{(0)}\ .
\end{equation} 
Note that $r_0r_c \ll 1$ implies that $\log(\frac{1}{r_0r_c})$ is
large which means that the $O(q^4)$ term need not be small even if we
are working withing the hydrodynamic regime \ie\
$\frac{q^2}{T^{2/z}}\ll 1$, suggesting a breakdown of the series
expansion. The expression for the the subleading part of $\chi$ \ie\
$\chi^{(1)}$ also changes to
\begin{equation}
\begin{split}
\chi^{(1)} \sim &\Big[\frac{q^2}{T^{2/z}}\Big\{(1-r_0r) -(1-r_0r)\log(1-r_0r)+(1-r_0r)\log^2(1-r_0r)\Big\}\Big.\\
+&\frac{q^4}{T^{4/z}}\Big\{-(1-r_0r)+(1-r_0r)\log(1-r_0r) -(1-r_0r)\log^2(1-r_0r)\Big.\\
+&\Big.\Big.(1-r_0r)\log^3(1-r_0r)\Big\}\log\Big(\frac{1}{r_0r_c}\Big)\Big]\tilde{\scriptA}_t^{(0)}+\frac{q^4}{T^{4/z}}\log^2(1-r_0r)\log^2\Big(\frac{1}{r_0r_c}\Big)\chi^{(0)}\ll \chi^{(0)}\,.
\end{split}
\end{equation}
From the preceding argument, we see again that the $O(q^4)$ term can
be arbitrarily large hinting at a breakdown of the series expansion.

\subsubsection*{Estimate for $h_{ty}^{(1)}$}

In the case when $z=4-\theta$ \eqref{series-eqn} takes the form
\begin{equation}
h_{ty}^{(0)}+h_{ty}^{(1)}+\cdots=\Big(\tilde{h}_{ty}^{(0)}+kr^{2-z}\int_{r_c}^{r}\frac{ds}{s}a_y^{(0)}\Big)+\Big(\tilde{h}_{ty}^{(1)}+kr^{2-z}\int_{r_c}^{r}\frac{ds}{s}a_y^{(1)}\Big)+\cdots\ .
\end{equation}
The above further implies that
\begin{equation}
\begin{split}
  h_{ty}^{(1)}&= \tilde{h}_{ty}^{(1)}+kr^{2-z}\int_{r_c}^{r}\frac{ds}{s}a_y^{(1)}\\
  &\sim \frac{q^4}{T^{4/z}}(1-r_0r)\log\left(\frac{1}{r_0r_c}\right)\left(1+\log \Big(\frac{1/r_0}{1/r_0-r}\Big)+\log^2\Big(\frac{1/r_0}{1/r_0-r}\Big) \right)\tilde{h}_{ty}^{(0)}\\
  &\hspace{4cm} +\frac{q^4}{T^{4/z}}(1-r_0r)\log^2 \Big(\frac{1}{r_0r_c}\Big)\log^2\Big(\frac{1/r_0}{1/r_0-r}\Big)\chi^{(0)}\ .
  \end{split}
\end{equation}
The above estimate is written using the estimate $\frac{\Gamma}{q}
\sim \frac{q}{T^{2/z}-1}\log(\frac{1}{r_0r_c})$. Further, the
assumptions \eqref{assumption-specialcase} implies that $h_{ty}^{(1)}$
may not be subleading compared to $h_{ty}^{(0)}$ thus suggesting a
breakdown of some sort in this analysis.

\section{Discussion}

In this paper, we have explored in greater detail our investigations
of shear diffusion in nonrelativistic hyperscaling violating Lifshitz
theories \cite{Kolekar:2016pnr}, adapting the membrane-paradigm-like
analysis \cite{Kovtun:2003wp} of near horizon perturbations. In
theories where a gauge field is present as a source for the
nonrelativistic metric (along with a scalar), some of the metric
perturbations $h_{ty}, h_{xy}$ mix with some of the gauge field
perturbations $a_y$. Since these are uncharged black branes, the
near-horizon region should still be characterized by simply
temperature and velocity variables, and charge cannot enter. Thus we
expect that the gauge field cannot dramatically alter the structure of
the near horizon diffusion equation found in \cite{Kolekar:2016pnr}
without the gauge field. Our analysis in this paper vindicates this:
we find a similar near-horizon analysis can be obtained resulting in a
diffusion equation for new field variables ${\tilde h_{xy}}\equiv
h_{xy}$ and ${\tilde h_{ty}}\equiv h_{ty}-r^{\theta-2}\int_{r_c}^r
s^{3-z-\theta}a_y ds$\ (for 4 bulk dimensions). Then, as in
\cite{Kolekar:2016pnr}, for $z<4-\theta$, we obtain universal
behaviour for the shear diffusion constant, suggesting that the
viscosity bound ${\eta\over s}={1\over 4\pi}$ holds. The regime
$z>4-\theta$ includes \eg\ hyperscaling violating theories arising
from the dimensional reduction of \eg\ $D6$-branes (giving $d_i=6,
z=1, \theta=9$) which do not admit a good gauge/gravity duality
(ill-defined asymptotics with gravity not decoupling): however it
might be interesting to find and understand reasonable holographic
theories whose exponents lie in this window. For $z=4-\theta$, we find
logarithmic behaviour as found previously. The hyperscaling violating
Lifshitz theories arising from $AdS$ plane waves (highly boosted black
branes) as well as nonconformal brane plane waves
\cite{Narayan:2012hk, Singh:2012un,Narayan:2013qga}, fall in this
category: this suggests that a null reduction of the hydrodynamics of
the boosted black brane might need a closer study to realize this in
detail, as we have described. We hope to explore this further.

We have seen the condition $z<2+d_i-\theta$\ (or $z<4-\theta$ here,
for bulk 4-dims) arising naturally from the perturbations falling off
asymptotically (\ref{tilde-At-integral}) in our case. We implicitly
regard hyperscaling violating theories as infrared phases arising from
\eg\ string realizations in the ultraviolet: however the window
$z<2+d_i-\theta$ ensures that the ultraviolet structure is essentially
unimportant, the diffusion constant arising solely from the near
horizon long-wavelength modes.  This still needs to be reconciled with
a clear holographic calculation: however some preliminary remarks are
as follows. We have seen that the ${\tilde h_{ty}}$ mode has
asymptotic fall-offs $r^{\theta-2} ({\tilde h_-}+\ldots) +
r^{2-z}({\tilde h_+}+\ldots)$ in bulk 4-dimensions.  For $z<4-\theta$,
the dominant mode near the boundary $r\ra r_c\sim 0$ is $r^{\theta-2}$
which is slower, leading to fixed $h_-$ boundary conditions relevant
for standard quantization ($h_-$ taken as source). This is the sector
that is continuously connected to $AdS$-like relativistic theories
($z=1, \theta=0$), as our perturbation analysis suggests. With the
conformal dimensions satisfying $\Delta_-+\Delta_+=2+z-\theta$
\cite{Dong:2012se} (see also
\cite{Chemissany:2014xsa,Taylor:2015glc}), the momentum density
operator ${\cal P}^i$ has dimension $3-\theta$: so taking
$\Delta_+=3-\theta$ gives $\Delta_-=z-1$ and $\Delta_-<\Delta_+$
implies $z<4-\theta$. In a reasonable theory where this is violated,
it would seem that the analog of alternative quantization
\cite{Klebanov:1999tb} is at work, with fixed $h_+$ boundary
conditions. In this light, $z=4-\theta$ is the case where the two
fall-offs coincide with $\Delta_-=\Delta_+$, and a logarithmic second
solution will arise suggesting logarithmic behaviour in the
correlation function as well. This is the case for $AdS$ plane waves 
(or highly boosted black branes): this may be interesting to explore.

It is worth putting the analysis here leading to (\ref{DT(2-z)/z}), 
(\ref{DT(2-z)/z:ay}), in perspective with the calculation of viscosity 
via the Kubo formula 
$\eta = -\lim_{\omega\ra 0} {1\over\omega} {\rm Im} G^R_{xy,xy}(\omega)$, 
with $G^R$ the retarded Green's function \cite{Son:2002sd}, assuming 
$T_{ij}\sim \eta (\del_iv_j + \ldots)$ in the dual field theory. 
The $h^x_y$ perturbation is modelled holographically as a massless 
scalar leading to the $\langle T_{xy}T_{xy}\rangle$ holographic 
correlation function (see \eg\ 
\cite{Pang:2009wa,Roychowdhury:2014lta,Kuang:2015mlf} for various 
subfamilies in (\ref{hvmetric})). For instance from 
\cite{Kuang:2015mlf}, the appropriate zero momentum ${\vec k}=0$ 
solutions to the scalar wave equation eventually lead to\
$G^R = -i{\omega\over 16\pi G} {R^{d_i}\over r_{hv}^\theta} r_0^{d_i-\theta}$ 
and thereby $\eta$: here the metric (\ref{hvmetric}) is written as\ 
$ds^2 = R^2 ({r\over r_{hv}})^{2\theta/d_i} (-f(r){dt^2\over r^{2z}} 
+ \ldots)$, retaining explicitly the dimensionful factors $R$ and 
the scale $r_{hv}$ inherent in these theories \cite{Dong:2012se}. 
Likewise the horizon area gives the entropy density\ 
$s={1\over 4G} {R^{d_i}\over r_{hv}^\theta} r_0^{d_i-\theta}$ which leads 
to ${\eta\over s}={1\over 4\pi}$ in agreement with our analysis. 
(We have seen that $\theta$ disappears from the temperature dependence 
of ${\cal D}$ in (\ref{DT(2-z)/z}): this is consistent with \eg\ 
cases where the hyperscaling violating phase arises from string 
constructions such as nonconformal branes which are known to have 
universal ${\eta\over s}$ behaviour.)

In light of the above, note that the Kubo analysis stemming from a
zero frequency $\omega\ra 0$ limit for the $h^x_y$ mode alone, does
not appear to give any insight into where a condition like
$z<2+d_i-\theta$ could arise from.  On the other hand, our analysis
here and in \cite{Kolekar:2016pnr} in terms of the near-horizon
perturbations involves the $h_{ty}$ perturbation as well (as in
\cite{Kovtun:2003wp}), which is coupled at nonzero $\omega$ to
$h^x_y$, and leads to the diffusion equation. The $h_{ty}$ mode (or
${\tilde h_{ty}}$ here) exhibits this nontrivial behaviour where the
normalizable mode can turn around depending on the exponents $z,
\theta$, the critical condition being the family $z=2+d_i-\theta$
where the two modes coincide.  This condition is trivially satisfied
for all relativistic theories of interest, with $z=1,\ \theta=0$, so
the Kubo limit is in perfect agreement with the near horizon diffusion
analysis. However in the present nonrelativistic cases, the near horizon
perturbations analysis appears to exhibit more structure. It would
seem that the structure of these perturbations is straightforward 
and simply involves analysing gravitational perturbations, not 
requiring detailed understanding of the holographic dictionary in 
this case. Therefore assuming that this is reliable, our analysis
suggests that the Kubo limit might need to be understood better in
theories where $z<2+d_i-\theta$ is violated. In the case with a gauge
field, the field variable ${\tilde h_{ty}}$ which exhibits this
behaviour naively suggests that perhaps a new energy-momentum tensor
variable ${\tilde T_{\mu\nu}}$ involving some linear combination of
$T_{\mu\nu}$ and the current density $j_\mu$ is the relevant
hydrodynamic observable that systematically encodes the
thermodynamic/hydrodynamic relations between the expansion of the
energy-momentum tensor, the shear viscosity $\eta$ and the diffusion
constant ${\cal D}$. We hope to explore these issues further.

\vspace{5mm}

{\footnotesize \noindent {\bf Acknowledgements:}\ \
It is a pleasure to thank Shiraz Minwalla for a useful conversation
during the course of this work. KN thanks the string theory group,
TIFR, Mumbai, for hospitality while this work was in progress. We 
thank the hospitality of the string group, IMSc, Chennai while this 
work was in progress, and the hospitality of the organizers of Indian 
Strings Meeting (ISM2016) as this work was being finalized. This work 
is partially supported by a grant to CMI from the Infosys Foundation.
}

\appendix
\section{Linearized equations for perturbations on $d=3$ 
hyperscaling violating background}\label{appendix-A}

In this section, we list the equations of motion for perturbations
$h_{tt}$, $h_{tx}$, $h_{xx}$, $h_{yy}$, $a_t$, $a_x$ and $\varphi$ for the 
sake of completeness. For $d=3$, the values of the various constants are
\begin{equation}
\beta\equiv \sqrt{(2-\theta)(2z-\theta-2)}\ , \quad \lambda=\frac{4-\theta}{\beta}\ , \quad \delta=\frac{\theta}{\beta}\ , \quad \Lambda=-\frac{1}{2}(2+z-\theta)(1+z-\theta)\ .
\end{equation}
The $t$, $x$ and $r$ components of the linearized Maxwell's equation
\eqref{lingaugeeqn}, respectively, give
\begin{equation}
\partial_r(r^{3+z-\theta}\partial_r a_t)-\frac{r^{3+z-\theta}}{f}(q^2 a_t+q\omega a_x)-\frac{k}{2}\Big[\partial_r(r^{2-\theta}(h_{xx}+h_{yy}))+\partial_r\Big(\frac{r^{2z-\theta}}{f}h_{tt}\Big)\Big]-k\lambda\partial_r\varphi=0\ ,
\end{equation}
\begin{equation}
\partial_r(r^{5-z-\theta}f\partial_r a_x)+\frac{r^{3+z-\theta}}{f}(q\omega a_t+\omega^2 a_x)-k\partial_r(r^{2-\theta}h_{tx})=0\ ,
\end{equation}
\begin{equation}
q[r^{5-z-\theta}f\partial_ra_x-kr^{2-\theta}h_{tx}]+\omega\Big[r^{3+z-\theta}\partial_r a_t-\frac{k}{2}\Big(r^{2-\theta}(h_{xx}+h_{yy})+\frac{r^{2z-\theta}}{f}h_{tt}\Big)-k\lambda\varphi\Big]=0\ .
\end{equation}
The $tt$-component of the linearized Einstein's equation
\eqref{lineinsteineqn} gives
\begin{equation}\begin{split}
& \partial_r^2 h_{tt}-\Big(\frac{2-4z+\theta}{r}+\frac{\partial_r f}{f}\Big)\frac{\partial_r h_{tt}}{2}-\frac{q^2}{f}h_{tt}-\frac{2q\omega}{f}h_{tx}-kr^{1-z}\partial_r a_t \\
& +\Big[\frac{-2(1+z-\theta)(2+z-\theta)}{r^{2}f}-\frac{2(2z-\theta)\partial_r f}{rf}+\frac{(\theta-2z)^{2}}{r^{2}}+\frac{(\partial_r f)^2}{f^2}\Big]\frac{h_{tt}}{2} \\
& -\frac{1}{2}\partial_r(r^{\theta-2z}f)\partial_r(r^{2-\theta}(h_{xx}+h_{yy})) -\frac{\omega^2}{f}(h_{xx}+h_{yy})+(2+z-\theta)\beta r^{-2-2z+\theta}\varphi=0\ .
\end{split}\end{equation}
The $tx$-component of \eqref{lineinsteineqn} gives
\begin{equation}
\partial_r(r^{z+\theta-3}\partial_r(r^{2-\theta}h_{tx}))+\frac{r^{z-1}}{f}q\omega h_{yy}-k\partial_r a_x=0\ .
\end{equation}
The $tr$-component of \eqref{lineinsteineqn} gives
\begin{equation}
q\partial_r\Big(\frac{r^{2z-\theta}}{f}h_{tx}\Big)+\frac{\omega}{2}\Big[\partial_r\Big(\frac{r^{2z-\theta}}{f}(h_{xx}+h_{yy})\Big)+\frac{r^{2z-2}}{f}\partial_r(r^{2-\theta}(h_{xx}+h_{yy}))\Big]+\omega\frac{r^{2z-3}}{f}\beta\varphi=0\ .
\end{equation}
Adding $xx$-component to $yy$-component of \eqref{lineinsteineqn} gives
\begin{equation}\begin{split}
& \partial_r(r^{2\theta-z-3}f\partial_r(r^{2-\theta}(h_{xx}+h_{yy})))+\omega^2\frac{r^{z+\theta-3}(h_{xx}+h_{yy})}{f}-2r^{\theta-z-1}q^2h_{yy}+\frac{2r^{z+\theta-3}}{f}q\omega h_{tx} \\
& -2kr^{\theta-2}\partial_r a_{t}-(\theta-2)r^{2\theta-z-4}f\partial_r\Big(\frac{r^{2z-\theta}h_{tt}}{f}\Big)+\frac{k^2 r^{z+\theta-5}}{f}h_{tt}+q^2\frac{r^{z+\theta-3}h_{tt}}{f} \\
& -\frac{2(2+z-\theta)(4-4z-2\theta+\theta^2)}{\beta}r^{2\theta-z-5}\varphi=0\ .
\end{split}\end{equation}
Subtracting $yy$-component from $xx$-component of \eqref{lineinsteineqn} gives
\begin{equation}
\partial_r(r^{\theta-z-1}f\partial_r(r^{2-\theta}(h_{xx}-h_{yy})))+\frac{r^{z-1}}{f}\omega^2(h_{xx}-h_{yy})+\frac{r^{z-1}}{f}q^2h_{tt}+\frac{2r^{z-1}}{f}q\omega h_{tx}=0\ .
\end{equation}
The $xr$-component of \eqref{lineinsteineqn} gives
\begin{equation}\begin{split}
& q[r^{2-\theta}\Big(\partial_r h_{tt}+\frac{1+z-\theta}{r}h_{tt}-\frac{\partial_r f}{2f}h_{tt}\Big)-r^{2-2z}f\partial_r(r^{2-\theta}h_{yy})-kr^{3-z-\theta}a_t-\beta r^{1-2z}f\varphi] \\
& +\omega[\partial_r(r^{2-\theta}h_{tx})-kr^{3-z-\theta}a_x]=0\ .
\end{split}\end{equation}
The $rr$-component of \eqref{lineinsteineqn} gives
\begin{equation}\begin{split}
& \partial_r^2(h_{xx}+h_{yy})+\Big(\frac{3(2-\theta)}{2r}+\frac{\partial_r f}{2f}\Big)\partial_r(h_{xx}+h_{yy})+(\theta-2)\Big(\frac{\theta}{2r^2}-\frac{\partial_r f}{2rf}\Big)(h_{xx}+h_{yy}) \\
& -\frac{r^{2z-2}}{f}\partial_r^2 h_{tt}+r^{2z-2}\partial_r h_{tt}\Big(\frac{-2-4z+3\theta}{2rf}+\frac{\partial_r f}{2f^2}\Big)+\alpha(2+z-\theta)\frac{r^{z-1}}{f}\partial_r a_{t} \\
& +\Big[\frac{\theta(2z-\theta)}{2r^2f}-\frac{(\partial_r f)^2}{2f^3}-\frac{1}{r^2f^2}\Big((z-1)(2+z-\theta)+(z-1)r\partial_r f-r^2\partial_r^2 f\Big)\Big]r^{2z-2}h_{tt}\\
&+2\beta r^{\theta-3}\partial_r\varphi- \frac{(2+z-\theta)\beta}{f}r^{\theta-4}\varphi=0\ .
\end{split}\end{equation}
The linearized scalar field equation \eqref{linscalareqn} gives
\begin{equation}\begin{split}
& \partial_r(r^{\theta-z-1}f\partial_r\varphi)+\Big(\frac{k^2\lambda^2}{2}-2\Lambda\delta^2\Big)r^{\theta-z-3}f\varphi+r^{\theta-z-1}\Big(\frac{r^{2z-2}\omega^2}{f^2}-\frac{q^2}{f} \Big)\varphi \\
& +\frac{\beta r^{\theta-z-2}f}{2}\Big[\partial_r(r^{2-\theta}(h_{xx}+h_{yy}))-\partial_r\Big(\frac{r^{2z-\theta}h_{tt}}{f}\Big)\Big]+\frac{k^2\lambda r^{z-3}h_{tt}}{2f}-k\lambda\partial_r a_{t}=0\ .
\end{split}\end{equation}

\section{Solutions to linearized equations for 
$h_{ty}$, $a_y$, $h_{xy}$ at zero momentum and zero 
frequency}\label{leading-soln}

At $q=0$, $\omega=0$, the linearized equations of motion 
\eqref{upstairseom1}-\eqref{upstairseom4} reduce to
\begin{eqnarray}
\label{upstairseom1zeroq}
\partial_r(r^{5-z-\theta}f\partial_r a_y)-\alpha(2+z-\theta)\partial_r(r^{2-\theta}h_{ty})&=&0\ ,\\
\label{upstairseom2zeroq}
\partial_r(r^{z+\theta-3}\partial_r(r^{2-\theta}h_{ty}))-\alpha(2+z-\theta)\partial_r a_y&=&0\ ,\\
\label{upstairseom3zeroq}
\partial_r(r^{-1-z+\theta}f\partial_r(r^{2-\theta}h_{xy}))&=&0\ .
\end{eqnarray}
For the sake of brevity, from now on we will denote the $\partial_r$
operator with a prime ``$\prime$'' on the functions.  Integrating
\eqref{upstairseom1zeroq} and substituting $\partial_r a_{y}$ in
\eqref{upstairseom2zeroq} gives
\begin{equation}\label{fhtyinhomog}
f(r)[r^{2}h_{ty}''+(1+z-\theta)rh_{ty}'-(\theta-2)(z-2)h_{ty}]-2(z-1)(2+z-\theta)h_{ty}=-\alpha^2(2+z-\theta)^2 c_1 r^{\theta-2}\ ,
\end{equation}
where we have chosen the integration constant as
$-\alpha(2+z-\theta)c_1$. This inhomogeneous equation has a particular
solution $h_{ty}=c_1r^{\theta-2}$. The homogeneous part of the above
equation,
\begin{equation}\label{fhtyhomog}
f(r)[r^{2}h_{ty}''+(1+z-\theta)rh_{ty}'-(\theta-2)(z-2)h_{ty}]-2(z-1)(2+z-\theta)h_{ty}=0\ 
\end{equation}
can be solved by substituting a series ansatz,
$h_{ty}=\sum_{n=0}^{\infty} c_{n}r^{m+n}$. Along with the two linearly
independent homogeneous solutions, the complete solution (including the
particular solution) is
\begin{equation}\label{fhtysoln}
h_{ty}=c_1 r^{\theta-2}+c_3r^{\theta-2z}f+c_4r^{z}\Big[1+\frac{(z-1)(r_{0}r)^{2+z-\theta}}{(1+2z-\theta)}\,_{2}F_{1}\Big(1,\frac{3z-\theta}{2+z-\theta},\frac{4+5z-3\theta}{2+z-\theta}; (r_{0}r)^{2+z-\theta} \Big) \Big]\,.
\end{equation}
Substituting $h_{ty}$ from the above expression in
\eqref{upstairseom2zeroq} and integrating, we get
\begin{equation}\begin{split}
a_{y}& =-\frac{C}{k} -\alpha c_3r^{-(2+z-\theta)} \\
& +c_4\Big[\frac{r^{2z-2}}{\alpha}+\alpha\frac{(2+z-\theta)}{(1+2z-\theta)}r_{0}^{2+z-\theta}r^{3z-\theta}\ _{2}F_{1}\Big(1,\frac{3z-\theta}{2+z-\theta},\frac{4+5z-3\theta}{2+z-\theta}; (r_{0}r)^{2+z-\theta} \Big) \Big. \\
& \Big. +\alpha\frac{r_{0}^{2+z-\theta}r^{1+3z-\theta}}{2(1+2z-\theta)}\ _{2}F^{\ '}_{1}\Big(1,\frac{3z-\theta}{2+z-\theta},\frac{4+5z-3\theta}{2+z-\theta}; (r_{0}r)^{2+z-\theta} \Big) \Big]\ ,
\end{split}\end{equation}
where $_{2}F^{\ '}_{1}=\frac{d}{dr}(_{2}F_{1})$. The last term \ie\ 
$_{2}F^{\ '}_{1}\left(1,\frac{3z-\theta}{2+z-\theta},\frac{4+5z-3\theta}{2+z-\theta}; (r_{0}r)^{2+z-\theta} \right)$ is in fact divergent at the horizon 
$r=\frac{1}{r_0}$. Integrating \eqref{upstairseom3zeroq}, we get
\begin{equation}
h_{xy}=b_{1}r^{\theta-2}\log(1-(r_0r)^{2+z-\theta})+b_{2}r^{\theta-2}\ .
\end{equation}

\section{Spatial compactification of the hyperscaling violating 
Lifshitz theory}\label{dim-redn}

The action \eqref{hvaction} in 4 bulk dimensions (\ie\ $d=3$) becomes
\begin{equation}
\label{hvaction-4d}
S= -\frac{1}{16 \pi G_{N}^{(4)}}\int d^4x \sqrt{-G}\left[R-\frac{1}{2}\partial_{\mu}\phi \partial^{\mu}\phi-\frac{Z(\phi)}{4}F_{\mu \nu}F^{\mu \nu}+V(\phi) \right]\ .
\end{equation}
In the $y$-compactified theory (where $y$ is one of the spatial
dimensions enjoying translation invariance), the metric component
$G_{yy}$ is parametrized by the scalar $\psi$ as $G_{yy}=e^{2\psi}$,
the gauge field coupling $Z =r^{4-\theta}$ and the other metric modes
are parametrized as
\begin{equation}
\label{compactifiedmetric}
G_{\mu \nu} \equiv
\begin{bmatrix}
\hat{g}_{ab}+e^{2\psi}\scriptA_{a}\scriptA_{b} & e^{2\psi}\scriptA_{a}\\
e^{2\psi}\scriptA_{b} & e^{2\psi}
\end{bmatrix}\ ,
\end{equation}
where $\hat{g}_{ab}$ is the metric of the compactified theory. Roman
indices $\{a,b\}$ runs over coordinates $t,r,x$ while Greek indices
$\{\mu, \nu\}$ runs over $t,r,x,y$. We also parametrize the gauge
field $A_{\mu}$ in the following way
\begin{equation}
A_{\mu} \equiv
\begin{bmatrix}
.\\
A_{a}\\
.\\
\chi
\end{bmatrix}\ .
\end{equation}
The gravity sector under compactification becomes
\begin{equation}
\begin{aligned}
S_{grav}&=-\frac{1}{16 \pi G_{N}^{(4)}}\int d^4x \sqrt{-G}\cdot R \\
&=-\frac{1}{16 \pi G^{(3)}}_{N}\int d^3x \cdot e^{\psi}\sqrt{-\hat{g}}\left(\hat{R}_{(3)}-\frac{1}{4}e^{4\psi}\scriptF_{ab}\scriptF^{ab}\right)\ ,
\end{aligned}
\end{equation}
where $\hat{R}_{(3)}$ is the Ricci scalar for the metric
$\hat{g}_{ab}$. The Maxwell action after a $y$-compactification can be
written as
\begin{equation}
\begin{aligned}
S_{Max}&=-\frac{1}{16 \pi G_N^{(4)}}\int d^{4}x \sqrt{-G}\left( -\frac{1}{4}ZF_{\mu \nu}F^{\mu \nu}\right)\\
&=-\frac{1}{16 \pi G_N^{(3)}}\int d^{3}x \left(-\frac{e^{\psi}Z\sqrt{\hat{g}}}{4}\right)\left[\hat{g}^{ac}\hat{g}^{bd}F_{ab}F_{cd}+4\hat{g}^{bc}F_{ab}\scriptA^{a}(\partial_c \chi)\right.\\
&\qquad \qquad \qquad \left. +2\hat{g}^{ab}(e^{-2\psi}+\scriptA_c \scriptA^c)(\partial_a \chi)(\partial_b \chi)-2\scriptA^{a}\scriptA^{b}(\partial_a \chi)(\partial_b \chi)\right]\ .
\end{aligned}
\end{equation}
A Weyl transformation $g_{ab}= e^{2\psi}\hat{g}_{ab}$ enables us to
write the gravitational and Maxwell sector of action after
compactification in the Einstein frame as
\begin{equation}\label{compax-action}
\begin{aligned}
S_{grav}+ S_{Max}&= -\frac{1}{16 \pi G_{N}^{(4)}}\int d^4x \sqrt{-G}\left[R-\frac{Z(\phi)}{4}F_{\mu \nu}F^{\mu \nu}\right]\\
&=-\frac{1}{16 \pi G^{(3)}_{N}}\int d^3x \cdot \sqrt{-g}\left(R_{(3)}-\frac{1}{4}e^{4\psi}\scriptF_{ab}\scriptF^{ab}\right.\\
&\qquad \quad +Ze^{2\psi}\left(-\frac{1}{4}F_{ab}F^{ab}-F_{a}^{\ c}\scriptA^{a}(\partial_c \chi)-\frac{1}{2}e^{-4\psi}(\partial_a \chi)(\partial^a \chi)\right.\\
&\qquad \quad \left.\left.-\frac{1}{2}\scriptA_c \scriptA^c(\partial_a \chi)(\partial^a \chi)+\frac{1}{2}\scriptA^a \scriptA^b (\partial_a \chi)(\partial_b \chi)\right)\right)\ ,
\end{aligned}
\end{equation}
where $R_{(3)}$ is the Ricci scalar of the 3-dimensional bulk metric
$g_{ab}$. The terms appearing in the last line of the above equation
will not contribute to the equations of motion at linearized order
since they appear at quartic order in the action. Varying the above
action w.r.t. the field $\scriptA_{\mu}$, at linearized level we get
\begin{equation}
\frac{1}{\sqrt{g}}\partial_a \left(\sqrt{-g}e^{4\psi}\scriptF^{ab}\right)=e^{2\psi}Zg^{ab}g^{cd}F_{ad}(\partial_c \chi)\ ,
\end{equation}
which for $b\equiv t,x,r$ gives \eqref{nu-t-eqn}, \eqref{nu-x-eqn} and
\eqref{nu-r-eqn} respectively.

\end{document}